\documentclass[reqno,11pt]{amsart}
\usepackage{amsmath,amssymb,amsthm,graphicx,a4wide,enumerate,url,tikz}
\usepackage[small,bf]{caption}
\usepackage{subcaption}
\usepackage{color}
\theoremstyle{plain}

\theoremstyle{definition}

\theoremstyle{remark}

\newcommand{\prn}[1]{\left(#1\right)}

\newcommand{\brk}[1]{\left[#1\right]}

\newcommand{\ud}[1]{\,\mathrm{d}#1}
\renewcommand{\vec}[1]{\mathbf{#1}}

\newcommand{\vv}{\mathrm{v}}
\newcommand{\vS}{\mathrm{v}_\text{S}} 
\newcommand{\rhoS}{\rho_\text{S}} 
\newcommand{\xS}{x_\text{S}}
\newcommand{\pr}{{\prime}}
\newcommand{\ppr}{{\prime\/\prime}}
\newcommand{\tvar}{t_*}

\definecolor{dgreen}{rgb}{0,0.5,0}

\begin{document}
\parskip.9ex

\title[Structural Properties of the Stability of Jamitons]
{Structural Properties of the Stability of Jamitons}
\author[R. Ramadan]{Rabie Ramadan}
\address[Rabie Ramadan]
{Department of Mathematics \\ Temple University \\ \newline
1805 North Broad Street \\ Philadelphia, PA 19122}
\email{rabie.ramadan@temple.edu}

\author[R. R. Rosales]{Rodolfo Ruben Rosales}
\address[Rodolfo Ruben Rosales]
{Department of Mathematics \\ Massachusetts Institute of Technology \\
77 Massachusetts Avenue \\ Cambridge, MA 02139}
\email{rrr@math.mit.edu}

\author[B. Seibold]{Benjamin Seibold}
\address[Benjamin Seibold]
{Department of Mathematics \\ Temple University \\ \newline
1805 North Broad Street \\ Philadelphia, PA 19122}
\email{seibold@temple.edu}
\urladdr{http://www.math.temple.edu/\~{}seibold}


\begin{abstract}
It is known that inhomogeneous second-order macroscopic traffic models can reproduce the phantom traffic jam phenomenon: whenever the sub-characteristic condition is violated, uniform traffic flow is unstable, and small perturbations grow into nonlinear traveling waves, called jamitons. In contrast, what is essentially unstudied is the question: which jamiton solutions are dynamically stable? To understand which stop-and-go traffic waves can arise through the dynamics of the model, this question is critical. This paper first presents a computational study demonstrating which types of jamitons do arise dynamically, and which do not. Then, a procedure is presented that characterizes the stability of jamitons. The study reveals that a critical component of this analysis is the proper treatment of the perturbations to the shocks, and of the neighborhood of the sonic points.
\end{abstract}

\maketitle

\section{Introduction}
\label{sec:introduction}
The modeling of vehicular traffic flow via mathematical equations is a key building block in traffic simulation, state estimation, and control. Important ways to describe traffic flow dynamics are microscopic/vehicle-based \cite{Pipes1953, Newell1961, BandoHesebeNakayama1995}, cellular \cite{NagelSchreckenberg1992, Daganzo1994}, and continuum models. This last class is the focus of this paper, particularly: inviscid macroscopic models \cite{LighthillWhitham1955, Richards1956, Underwood1961, Payne1971, Payne1979, Lebacque1993, AwRascle2000} that describe the spatio-temporal evolution of the vehicle density (and other field quantities) via hyperbolic conservation laws. Other types of continuum models exist as well, including gas-kinetic \cite{HermanPrigogine1971, Phillips1979, IllnerKlarMaterne2003}, dispersive \cite{KurtzeHong1995, KomatsuSasa1995}, and viscous \cite{KernerKonhauser1993, KernerKonhauser1994} models. Hyperbolic models do not resolve zones of strong braking, but rather approximate them by traveling discontinuities (shocks) whose dynamics are described by appropriate jump conditions \cite{Evans1998}.
Macroscopic models play a central role in traffic flow theory and practice because:
\begin{itemize}
\item
Mathematically, other types of descriptions reduce/converge to macroscopic models in certain limits, including: microscopic \cite{AwKlarMaterneRascle2002}, cellular \cite{AlperovichSopasakis2008}, and gas-kinetic \cite{IllnerKlarMaterne2003, AlperovichSopasakis2008}.
\item
Practically, macroscopic models are best-suited for state estimation \cite{WangPapageorgiou2005, WorkTossavainenBlandinBayenIwuchukwuTracton2008}, for incorporating sparse GPS data \cite{Aminetal2008, HerreraWorkBanHerringJacobsonBayen2010}, and for control \cite{Papageorgiou1998}.
\item
Computationally, a macroscopic description is a natural framework to upscale millions of vehicles to a cell-transmission model \cite{Daganzo1994} with much fewer degrees of freedom.
\item
Societally, traffic descriptions that do not resolve individual vehicles are desirable for privacy and data security.
\end{itemize}

In this work, we focus on the lane-aggregated description of traffic flow dynamics on uniform highways without any road variations, let alone intersections or bottlenecks. The reason is that even in this simple scenario, real traffic flow tends to develop complex nonlinear dynamics, particularly the \emph{phantom traffic jam} phenomenon \cite{Kerner2000a, Helbing2001}: initially uniform flow develops (under small perturbations) into nonlinear traveling waves, called \emph{jamitons} \cite{FlynnKasimovNaveRosalesSeibold2009}. This occurrence of instabilities and waves without discernible reason has been demonstrated and reproduced experimentally \cite{SugiyamaFukuiKikuchiHasebeNakayamaNishinariTadakiYukawa2008, SternCuiDelleMonacheBhadaniBuntingChurchillHamiltonHaulcyPohlmannWuPiccoliSeiboldSprinkleWork2018}. While these features can be reproduced in microscopic car-following models, a key goal is to capture these non-equilibrium phenomena via macroscopic models (to facilitate the model advantages described above).

The archetype macroscopic model is the \emph{Lighthill-Whitham-Richards (LWR) model} \cite{LighthillWhitham1955, Richards1956}
\begin{equation}
\label{eq:LWR}
\rho_t+Q(\rho)_x = 0\;,
\end{equation}
that describes the evolution of the vehicle density $\rho(x,t)$ where $x$ is the road position and $t$ is time. The \emph{fundamental diagram} (FD) function $Q(\rho) = \rho U(\rho)$, where the \emph{equilibrium velocity function} $U(\rho)$ is the bulk flow velocity as a function of density, is motivated by the 1935 measurements by Greenshields \cite{Greenshields1935}, and many types of FD have been proposed \cite{LighthillWhitham1955, Greenberg1959, Underwood1961, Newell1993, Daganzo1994}. As a matter of fact, real FD data exhibits a substantial spread in the congested regime \cite{Kerner2000a}. More complex traffic models capture this spread \cite{Colombo2002, SeiboldFlynnKasimovRosales2013, FanSeibold2013, FanHertySeibold2014}, but the LWR model does not. Yet, due to its simplicity it nevertheless is widely used. Moreover, as we highlight below, it also is motivated as a reduced equation for more complex models.

Another critical shortcoming of the LWR model is that it cannot reproduce the phantom traffic jam phenomenon: being a \emph{first-order model}, it exhibits a maximum principle, and thus small perturbations to a uniform solution cannot amplify (instead, they turn into N-waves and decay). In this work, we focus on \emph{second-order models} that augment the vehicle density $\rho(x,t)$ by an independent field variable for the bulk velocity $u(x,t)$, and describe their evolution via a $2\times 2$ balance law system, specifically: a hyperbolic conservation law system with a relaxation term in the velocity equation. Due to conservation of vehicles, the density always evolves by the continuity equation, $\rho_t + (\rho u)_x = 0$. In turn, the velocity equation encodes the actual modeling of the vehicle dynamics and interactions. The \emph{Payne-Whitham (PW) model} \cite{Payne1971, Whitham1974}
\begin{equation}
\label{eq:PW_model}
\begin{split}
\rho_t+(\rho u)_x &= 0\;,\\
u_t + uu_x + p(\rho)_x/\rho &= \tfrac{1}{\tau}(U(\rho)-u)
\end{split}
\end{equation}
was the first second-order model proposed. Here $U(\rho)$ is the \emph{desired velocity function}, and $\tau$ is the relaxation time that determines how fast drivers adjust to their desired velocity $U(\rho)$. The \emph{traffic pressure} $p(\rho)$ models preventive driving. Even though the PW model does capture traffic waves accurately \cite{FlynnKasimovNaveRosalesSeibold2009, SeiboldFlynnKasimovRosales2013}, it is generally rejected \cite{Daganzo1995} due to spurious shocks that overtake vehicles from behind; and other hyperbolic models are preferred (see below). However, the fundamental structure of a $2\times 2$ hyperbolic system with a relaxation in the second equation, is common to all models of interest in this study.

Models with the structure described above possess a critical phase transition. If the \emph{sub-characteristic condition} (SCC) is satisfied, then uniform flow is stable \cite{Whitham1959, Whitham1974, Liu1987, ChenLevermoreLiu1994}. Conversely, when it is violated, uniform flow is unstable and nonlinear traveling wave solutions exist \cite{Li2000, JinKatsoulakis2000, Noble2007a, FlynnKasimovNaveRosalesSeibold2009, SeiboldFlynnKasimovRosales2013}. The SCC is defined as follows. Let $\lambda_1 < \lambda_2$ be the two characteristic speeds of the hyperbolic part of the model, and let $\mu = Q'(\rho)$ be the characteristic speed of the \emph{reduced equation} \eqref{eq:LWR} (with $Q(\rho) = \rho U(\rho)$), which arises in the formal limit $\tau\to 0$; in which $u$ relaxes infinitely fast to $U(\rho)$. Then the SCC is: $\lambda_1 \le \mu \le \lambda_2$.

The case of the SCC satisfied is well studied \cite{Whitham1959, Whitham1974, Liu1987, ChenLevermoreLiu1994, LiLiu2005}. In particular, it is related to positive diffusion when conducting a Chapman-Enskog expansion of the model \cite{KurtzeHong1995, HelbingJohansson2009}. In contrast, this paper focuses on understanding the behavior and stability of solutions when the SCC is violated.

This paper is organized as follows. In \S\ref{sec:models}, we introduce the equations. Then we characterize the nature of the instabilities to uniform flow, and the traveling wave solutions that then arise: the jamitons. In \S\ref{sec:computational_study}, a systematic computational study of the stability of jamitons is conducted. Those results then motivate a stability analysis of those nonlinear traveling waves, presented in \S\ref{sec:jamiton_stability_analysis}. We close with a discussion and a broader outlook in \S\ref{sec:conclusions}.

\section{Macroscopic Traffic Models with Instabilities and Traveling Waves}
\label{sec:models}
While the general results and methodologies apply to a wide class of second-order models with relaxation (including the PW model \eqref{eq:PW_model} and generic second-order models \cite{LebacqueMammarHajSalem2007, FanHertySeibold2014}), we focus this study on the inhomogeneous \emph{Aw-Rascle-Zhang (ARZ) model} \cite{AwRascle2000, Zhang2002}. In non-conservative form it reads as
\begin{equation}
\label{eq:ARZ_model}
\begin{split}
\rho_t+(\rho u)_x &= 0\;,\\
(u+h(\rho))_t + u(u+h(\rho))_x &= \tfrac{1}{\tau}(U(\rho)-u)\;,
\end{split}
\end{equation}
where $h(\rho)$ is called the \emph{hesitation function}. We assume that: $U(\rho)$ is strictly decreasing, $Q(\rho) = \rho U(\rho)$ is strictly concave, $h(\rho)$ is strictly increasing, and $\rho h(\rho)$ is strictly convex. In particular these assumptions yield a hyperbolic system, which has no waves that overtake vehicles (the 2-waves are contacts) \cite{AwRascle2000}. While originally proposed in homogeneous form, the addition of the relaxation term \cite{Greenberg2001} allows for the violation of the SCC.

In the homogeneous ARZ model, the field $w = u+h(\rho)$ can be interpreted as a convected quantity moving with the flow (the hesitation function reduces the empty road velocity $w$ by $h(\rho)$). Hence, the conserved variables are $\rho$ and $q = \rho(u+h(\rho))$, and the conservative form of the equations is
\begin{equation}
\label{eq:ARZ_model_conservative}
\begin{split}
\rho_t+\prn{q-\rho h(\rho)}_x &= 0\;,\\
q_t + \prn{\tfrac{q^2}{\rho}-q h(\rho)}_x &= \tfrac{1}{\tau}\prn{\rho (U(\rho)+h(\rho))-q}\;,
\end{split}
\end{equation}
with associated Rankine-Hugoniot jump conditions
\begin{equation}
\label{eq:ARZ_RH_jump_conditions}
\begin{split}
s\brk{\rho} - \brk{\rho u} &= 0\;,\\
s\brk{\rho\big(u+h(\rho)\big)} - \brk{\rho u^2 +\rho u h(\rho)} &= 0\;.
\end{split}
\end{equation}
Here $\brk{\zeta}$ denotes the jump of the variable $\zeta$ across the discontinuity, and $s$ is the speed. In addition, the Lax entropy conditions \cite{Evans1998} must be satisfied. Specifically: one family of characteristics goes through the discontinuity, while the other converges into it (for a shock), or is parallel to it (for a contact). In particular, the assumptions on $h$ made below \eqref{eq:ARZ_model} guarantee that the entropy conditions are equivalent to: the shocks are compressive (i.e., as vehicles go through a shock, the density increases) and move slower than the vehicles \cite{SeiboldFlynnKasimovRosales2013}.

The characteristic speeds of \eqref{eq:ARZ_model_conservative} are:
\begin{equation}
\label{eq:eigenvalues_ARZ}
\lambda_1 = q/\rho - h(\rho) - \rho h'(\rho) = u-\rho h'(\rho)\;,
\quad\text{and}\quad \lambda_2 = q/\rho - h(\rho) = u\;,
\end{equation}
where the $\lambda_1$ is genuinely nonlinear (associated with shocks and rarefactions), while the $\lambda_2$ is linearly degenerate (associated with contacts).

%
%

\subsection{Specific model functions}
\label{subsec:specific_model}
While the analysis and general results derived below hold for generic models \eqref{eq:ARZ_model_conservative}, the computational study and the illustrative graphs are presented for a specific choice of model functions. As in \cite{SeiboldFlynnKasimovRosales2013}, we choose $\rho_\text{max} = 1/7.5\text{m}$, $u_\text{max} = 20\text{m}/\text{s}$, and construct the fundamental diagram function
\begin{equation*}
Q(\rho) = c \prn{ g(0) + \prn{g(1)-g(0)} \tfrac{\rho}{\rho_\text{max}} -g\prn{\tfrac{\rho}{\rho_\text{max}}} }\;,
\quad\text{where}\ \
g(y) = \sqrt{1+\prn{\tfrac{y-b}{\lambda}}^2}\;,
\end{equation*}
that is a smoothed version of the Newell-Daganzo triangular flux \cite{Newell1993, Daganzo1994}. The parameters are chosen $c = 0.078 \rho_\text{max} u_\text{max}$, $b=\frac{1}{3}$, and $\lambda = \frac{1}{10}$ to have the function fit real sensor data \cite{SeiboldFlynnKasimovRosales2013}. Hence $U(\rho) = Q(\rho)/\rho$. Moreover, we choose $h(\rho) = 8\text{m}/\text{s}\sqrt{\frac{\rho}{\rho_\text{max} - \rho}}$, and the relaxation time $\tau=3$s. Note that these values are for a single lane. When considering multi-lane traffic, realistic values result by scaling $\rho$ and $Q$ by the number of lanes.

\subsection{Linear stability of uniform flow}
\label{subsec:stability_uniform_flow}
Before analyzing the stability of nonlinear waves, we discuss important aspects regarding the stability of uniform flow, i.e., base state solutions of \eqref{eq:ARZ_model} in which $\rho = \tilde{\rho}$ and $u = U(\tilde{\rho})$ are constant in space and time. The linear stability analysis itself is a well-established normal models analysis \cite{KernerKonhauser1993, FlynnKasimovNaveRosalesSeibold2009}, 
and we briefly outline the key steps. Consider infinitesimal wave perturbations (where $k$ is the wave number and $\sigma$ the complex growth rate) of the base state\;,
\begin{equation*}
\hat{\rho} = \hat{R} e^{ikx+\sigma t}
\quad\text{and}\quad
\hat{u} = \hat{U} e^{ikx+\sigma t}\;,
\end{equation*}
substitute the perturbed solution $\rho = \tilde{\rho}+\hat{\rho}$ and $u = U(\tilde{\rho})+\hat{u}$ into \eqref{eq:ARZ_model}, and consider only constant and linear terms. This leads to the system
\begin{equation}
\label{eq:linear_stability_system}
\begin{bmatrix}
\sigma +ik\psi & ik\tilde{\rho}\\
\sigma \phi + ik\psi\phi -\frac{\xi}{\tau} & \sigma +ik\psi +\frac{1}{\tau}
\end{bmatrix}
\begin{bmatrix} \hat{R}\\ \hat{U}
\end{bmatrix} =
\begin{bmatrix} 0\\ 0 \end{bmatrix},
\end{equation}
for the perturbation amplitudes, where $\psi = U(\tilde{\rho}) > 0$, $\phi = h'(\tilde{\rho}) > 0$, and $\xi = U'(\tilde{\rho}) < 0$. Nontrivial solutions can only exist if the matrix in \eqref{eq:linear_stability_system} has vanishing determinant, which requires
\begin{equation*}
\sigma = -ik\psi + ik \tfrac{1}{2}\tilde{\rho}\phi - \tfrac{1}{2\tau}(1+\Gamma)\;,
\end{equation*}
where $\Gamma$ satisfies $\Gamma^2 = 1-k^2\tau^2\tilde{\rho}^2\phi^2 - 2ik\tau\tilde{\rho}(\phi+2\xi)$. Writing $\Gamma = \Lambda_1+i\Lambda_2$ in terms of its real and imaginary part yields the two equations $\Lambda_1^2-\Lambda_2^2 = 1-k^2\tau^2\tilde{\rho}^2\phi^2$ and $\Lambda_1\Lambda_2 = k\tau\tilde{\rho}(\phi+2\xi)$, which then leads to the following quadratic equations for $z = (\Lambda_1)^2$:
\begin{equation}
\label{eq:quadratic_equation_z}
z^2 - (1-\beta^2k^2)z - \gamma^2k^2 = 0\;.
\end{equation}
Here $\beta = \tau\tilde{\rho}\phi$ and $\gamma = \tau\tilde{\rho}(\phi+2\xi)$. The positive solution of \eqref{eq:quadratic_equation_z}, as a function of $k$, is
\begin{align}
\label{eq:formula_z_ver1}
z^+(k) &= \tfrac{1}{2}\left((1-\beta^2k^2)+\sqrt{(1-\beta^2k^2)^2+4\gamma^2k^2}\right)\\
&= \tfrac{1}{2}\left((1-\beta^2k^2)+\sqrt{(1+\beta^2k^2)^2+4(\gamma^2-\beta^2)k^2}\right)\;.
\label{eq:formula_z_ver2}
\end{align}
This function has the following properties:
\begin{enumerate}[(i)]
\item $z^+(0) = 1$.
\item $\lim_{k\to\infty} z^+(k) = (\gamma/\beta)^2$, which follows from \eqref{eq:formula_z_ver1} and the asymptotic ($k\gg 1$) formula:
\begin{equation*}
\sqrt{(1-\beta^2k^2)^2+4\gamma^2k^2}
\sim
\beta^2k^2\sqrt{1+2(2\gamma^2-\beta^2)\beta^{-4}k^{-2}}
\sim
\beta^2k^2+(2(\gamma/\beta)^2-1)\;.
\end{equation*}
\item It is strictly monotonic if $|\gamma|\neq |\beta|$, i.e., it is strictly increasing if $|\gamma|>|\beta|$ and strictly decreasing if $|\gamma|<|\beta|$. This fact follows from \eqref{eq:formula_z_ver1}, because the sign of the term $4(\gamma^2-\beta^2)k^2$ determines the slope of $z^+(k)$: if $|\gamma|=|\beta|$, it is constant; and if the term is positive (negative), the function goes up (down) with $k$.
\end{enumerate}
The growth rate of normal modes is
\begin{equation*}
g_{\tilde{\rho}}(k) = \text{Re}(\sigma) = -\tfrac{1}{2\tau}(1+\text{Re}(\Gamma)) = -\tfrac{1}{2\tau}(1+\Lambda_1) = -\tfrac{1}{2\tau}\!\left(1\pm\sqrt{z^+(k)}\right)\;.
\end{equation*}
Linear stability, i.e., $\text{Re}(\sigma)\le 0$, is equivalent to $z^+\le 1$ (only the negative root of $\sqrt{z^+}$ could cause positive growth). Hence, stability holds exactly if $|\gamma|<|\beta|$, or equivalently $\phi+\xi>0$, or equivalently
\begin{equation}
\label{eq:linear_stability_condition}
h'(\tilde{\rho}) + U'(\tilde{\rho}) \ge 0\;.
\end{equation}
This last condition is exactly what the sub-characteristic condition (SCC) \cite{Whitham1959, Whitham1974} yields as well \cite{SeiboldFlynnKasimovRosales2013}: the LWR characteristic speed, $\mu = Q'(\tilde{\rho}) = U(\tilde{\rho})+\tilde{\rho} U'(\tilde{\rho})$ lies in between the two ARZ characteristic speeds, $\lambda_1 = U(\tilde{\rho})-\tilde{\rho} h'(\tilde{\rho})$ and $\lambda_2 = U(\tilde{\rho})$, exactly if \eqref{eq:linear_stability_condition} holds.


To recap, for the inhomogeneous ARZ model \eqref{eq:ARZ_model}, there are exactly two possibilities: Either the stability condition (the SCC) \eqref{eq:linear_stability_condition} holds; then all basic wave perturbations $e^{ikx}$ have non-positive growth rates, and solutions are linearly stable. Or \eqref{eq:linear_stability_condition} is violated; then all waves grow. Moreover, the rate of growth $g_{\tilde{\rho}}(k)$ is an increasing function of the wave number $k$, that has $g_{\tilde{\rho}}(0) = 0$, and approaches (as $k\to\infty$) the asymptotic growth rate \begin{equation*}
g_{\tilde{\rho}}^\infty = \lim_{k\to\infty} g_{\tilde{\rho}}(k)
= \tfrac{1}{2\tau}\!\left(|\gamma/\beta| - 1\right)
= \tfrac{1}{2\tau}\!\left( |1+2\,\xi/\phi| - 1\right)
= \tfrac{1}{\tau}\!\left( \tfrac{-U'(\tilde{\rho})}{h'(\tilde{\rho})} - 1\right).
\end{equation*}

\begin{figure}
\begin{subfigure}{.33\textwidth}
  \centering\captionsetup{width=.9\linewidth}%
  \includegraphics[width=.999\linewidth]{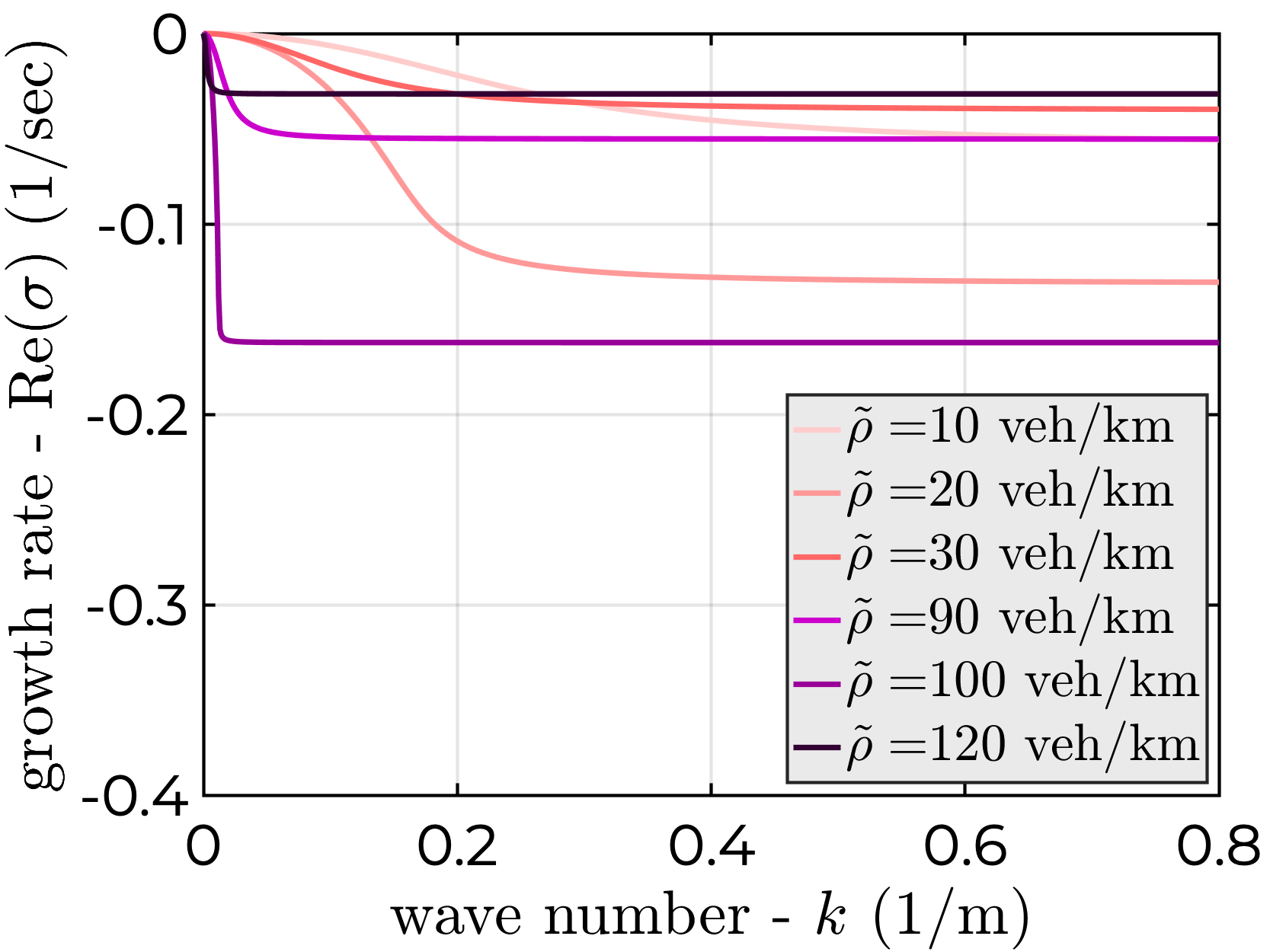}
  \caption{Growth rates $g_{\tilde{\rho}}(k)$ for different $\tilde{\rho}$ that satisfy \eqref{eq:linear_stability_condition}, i.e., are linearly stable.}
  \label{subfig:growth_rate_stable}
\end{subfigure}%
\begin{subfigure}{.33\textwidth}
  \centering\captionsetup{width=.9\linewidth}%
  \includegraphics[width=.999\linewidth]{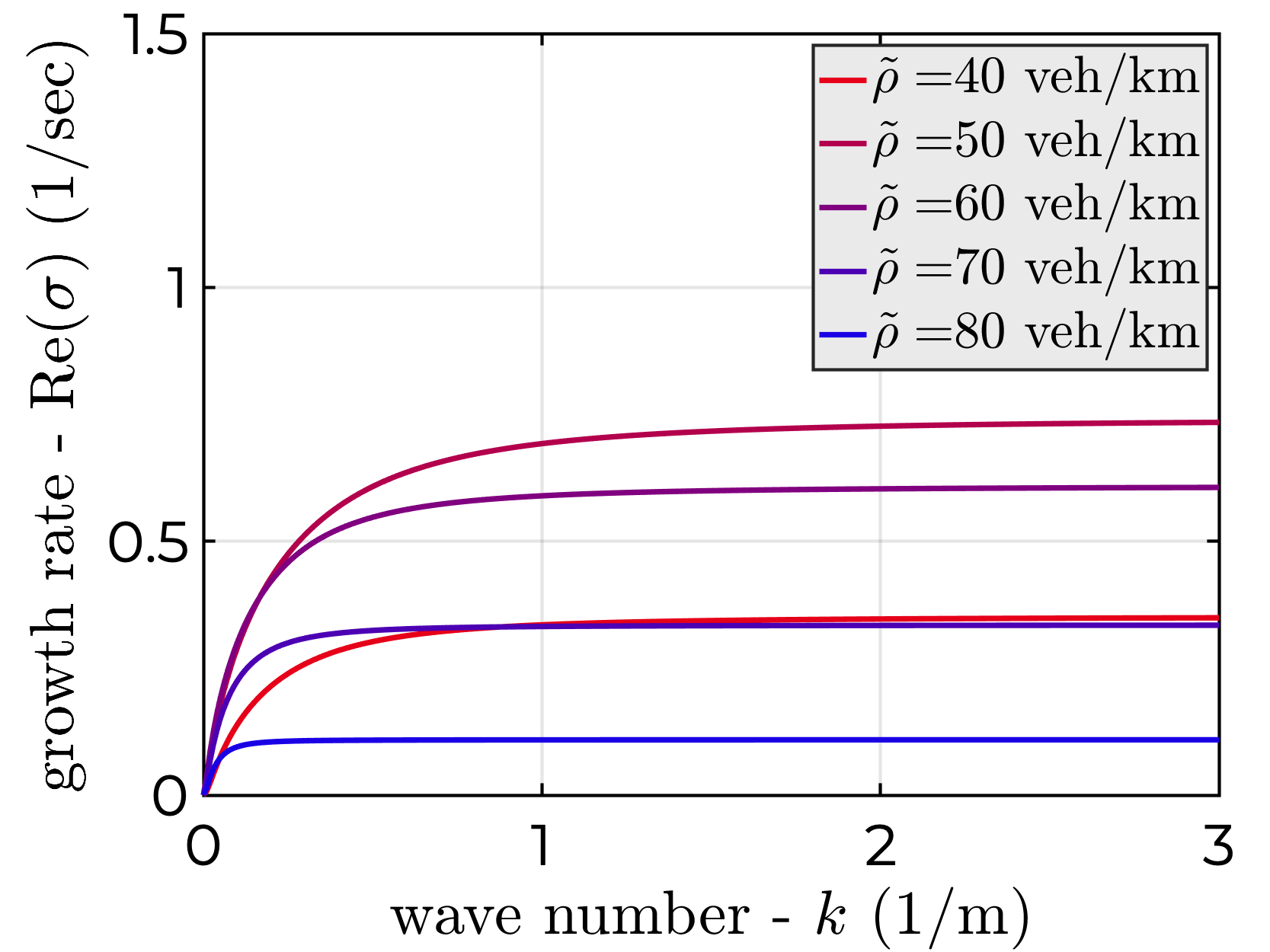}
  \caption{Growth rates $g_{\tilde{\rho}}(k)$ for different $\tilde{\rho}$ that violate \eqref{eq:linear_stability_condition}, i.e., are linearly unstable.}
  \label{subfig:growth_rate_unstable}
\end{subfigure}%
\begin{subfigure}{.33\textwidth}
  \centering\captionsetup{width=.9\linewidth}%
  \includegraphics[width=.999\linewidth]{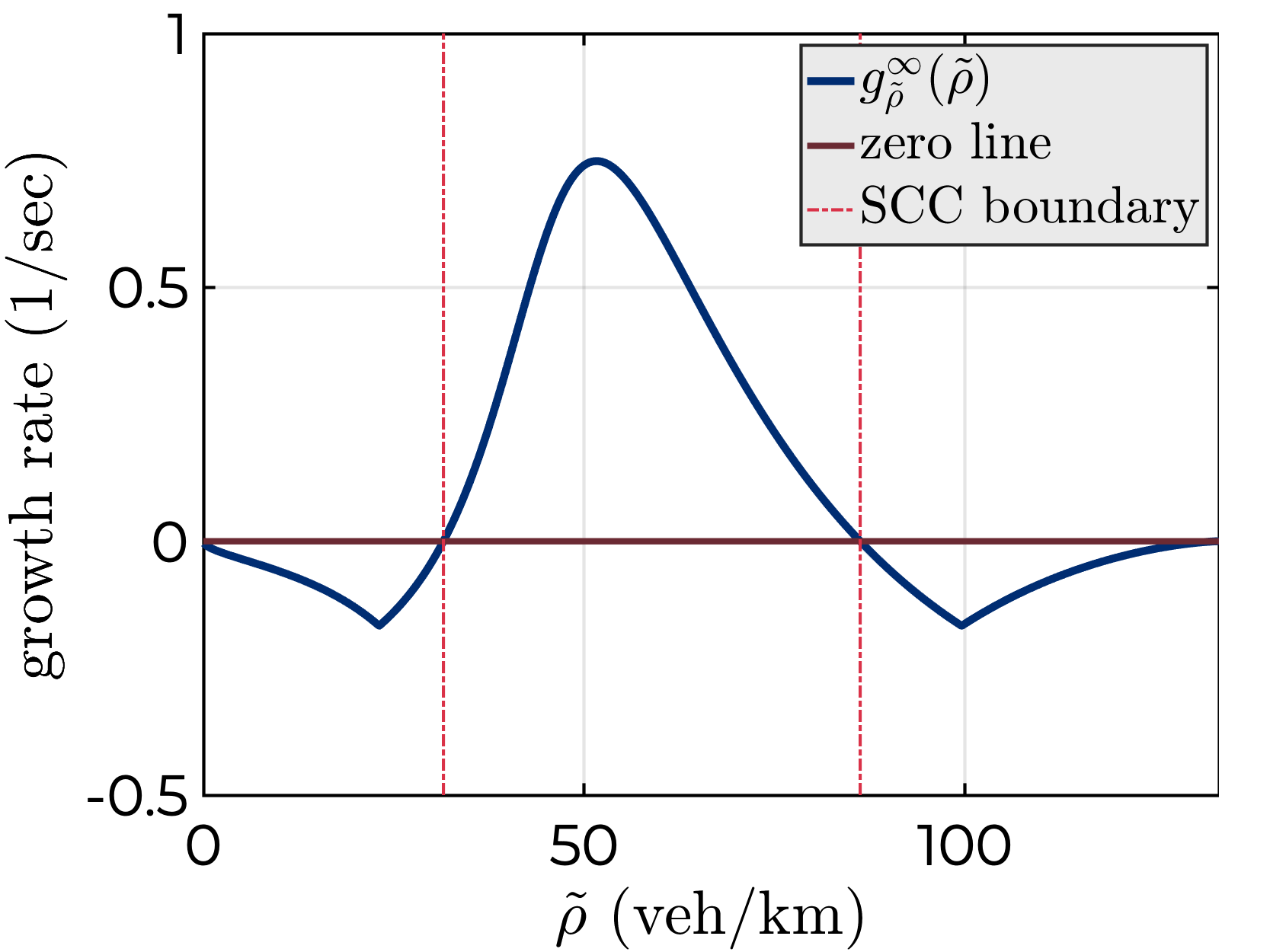}
  \caption{Asymptotic growth rate (worst case) $g_{\tilde{\rho}}^\infty = \lim_{k\to\infty} g_{\tilde{\rho}}(k)$ as a function of $\tilde{\rho}$.}
  \label{subfig:growth_rate_limit}
\end{subfigure}%
\caption{Plots of the growth rate $g_{\tilde{\rho}}(k) = \operatorname{Re}(\sigma)$ as a function of the wave number $k$, for different constant base states $\tilde{\rho}$, as well as the asymptotic growth rate $g_{\tilde{\rho}}^\infty$ as a function of $\tilde{\rho}$.}
\label{fig:growth_rate}
\end{figure}

Figure~\ref{fig:growth_rate} shows the growth rate functions $g_{\tilde{\rho}}(k)$ for the specific model given in \S\ref{subsec:specific_model}, with stable base states in panel~\ref{subfig:growth_rate_stable} and unstable base states in panel~\ref{subfig:growth_rate_unstable}. In the latter, one can clearly see the strict increase of $g_{\tilde{\rho}}$ with $k$, and the asymptotic limit $g_{\tilde{\rho}}^\infty$. Panel~\ref{subfig:growth_rate_limit} shows a plot of the asymptotic growth rate $g_{\tilde{\rho}}^\infty$ as a function of $\tilde{\rho}$.

Clearly, base states that satisfy \eqref{eq:linear_stability_condition} are well-behaved. However, with regards to modeling phantom traffic jams and jamitons, we are particularly interested in base states that violate \eqref{eq:linear_stability_condition}. These require some more careful discussion. While instabilities to uniform states are ubiquitous in science and engineering, having a growth rate that is increasing for all wave numbers is unusual. The much more common scenario (for example, fluid instabilities moderated by viscosity or surface tension \cite{DrazinReid1981}) is that medium wave length are unstable and short waves (i.e., $k$ large) are stable again, yielding a critical wave number $k^*$ of maximal growth. In that case, one can argue that out of infinitesimal perturbations, in which all wave lengths are present, the linearized dynamics will single out the ones with dominant growth. Hence, the wave number $k^*$ will be selected to first enter the nonlinear regime.

However, arguments of that type do not work for \eqref{eq:ARZ_model} because, as we have shown, its growth function $g_{\tilde{\rho}}(k)$ has no maximum. Rather, the shorter the waves in the perturbation, the faster their growth. It should be stressed that despite this behavior, the linearized model for \eqref{eq:ARZ_model} is mathematically well-posed: for any final time $t$, the amplification of normal modes is bounded by $\exp(t\,g_{\tilde{\rho}}^\infty)$. Still, from an application perspective, properly answering the question of which wave lengths dominate once an amplified perturbation leaves the linear regime, is important; but it is more challenging than in the usual situation.

While the PDE model \eqref{eq:ARZ_model} has no maximum wave number, reality does, namely the vehicle scale. Specifically, wave numbers beyond a $k_\text{max}$, given by the minimum spacing between vehicles, have no practical meaning. One possible way to exclude features on such unphysically short length scales is to add a small amount of viscosity to the ARZ model \eqref{eq:ARZ_model}, as in Kerner-Konh{\"a}user \cite{KernerKonhauser1993, KernerKonhauser1994} for the PW model \eqref{eq:PW_model}. In Fig.~\ref{subfig:growth_rate_unstable}, this would change the functions $g_{\tilde{\rho}}(k)$ to drop off once $k$ gets close to the vehicle scale. Similarly, the numerical discretization of the PDE \eqref{eq:ARZ_model} on grids that are never finer than the vehicle scale will produce a wave number cut-off via numerical viscosity of the method \cite{LeVeque1992}.

Another possibility (employed here in \S\ref{sec:computational_study}) is to consider small perturbations, rather than infinitesimal perturbations, and provide a model for the noise. Specifically, we argue that on real roads, perturbations of all wave lengths $k\in [0,k_\text{max}]$ will act: $k<k_\text{max}$ due to small variations in road features, wind, etc.; and $k\approx k_\text{max}$ due to variabilities across vehicles. The simplest such noise model is one where all wave numbers $k\in [0,k_\text{max}]$ appear with equal amplitudes, and perturbations with $k > k_\text{max}$ do not occur.

Because the growth function tends to have a plateau near $k_\text{max}$ (see Fig.~\ref{fig:growth_rate}), this linear growth/noise model will yield that all wave numbers $k$ near but below $k_\text{max}$ will be amplified to reach the nonlinear regime at the same time. This is not unrealistic, as it means that noise close to the vehicle scale will dominate before systematic nonlinear wave effects kick in.

As a final remark we wish to point out that once solutions of the ARZ model \eqref{eq:ARZ_model} leave the linear regime (around a uniform base state), the nonlinear dynamics tend to turn those vehicle-scale waves into oscillations with shocks that then collide and merge to form nonlinear wave structures of much smaller amplitude to wave-length ratios. However, those nonlinear transient dynamics are extremely complicated, and this insight is merely based on our observations from numerous highly resolved computations (like those done in \S\ref{sec:computational_study}). What we \emph{will} study, though, is the stability of true traveling wave solutions of \eqref{eq:ARZ_model} (jamitons) in the situation when the SCC \eqref{eq:linear_stability_condition} is violated (see \S\ref{sec:jamiton_stability_analysis}).

\subsection{Traveling wave analysis and jamitons}
\label{subsec:jamiton_construction}
Before studying waves, it is important to stress that macroscopic models (without explicit lane changing) can equivalently be written in Lagrangian variables. In \eqref{eq:ARZ_model_conservative} the equations are cast in Eulerian variables $\rho(x,t)$ and $q(x,t)$. The Lagrangian formulation, as used in \cite{Greenberg2001, SeiboldFlynnKasimovRosales2013}, employs the variables $\vv(\sigma,t)$ and $u(\sigma,t)$, where $\sigma$ is the (continuous) vehicle number, defined so that $\ud{\sigma} = \rho\ud{x}-\rho u\ud{t}$, and $\vv = 1/\rho$ is the specific traffic volume, i.e., the road length per vehicle. In these variables the ARZ model reads as
\begin{equation}
\label{eq:ARZ_model_lagrangian}
\begin{split}
\vv_t - u_\sigma &= 0\;,\\
(u+\hat{h}(\vv))_t &= \tfrac{1}{\tau} (\hat{U}(\vv)-u)\;,
\end{split}
\end{equation}
where $\hat{h}(\vv) = h(1/\vv)$ and $\hat{U}(\vv) = U(1/\vv)$. The assumptions on the model functions in Eulerian variables ($\frac{\ud{U}}{\ud{\rho}}<0$, $\frac{\ud{}^2Q}{\ud{\rho^2}}<0$, $\frac{\ud{h}}{\ud{\rho}}>0$, $\frac{\ud{}^2}{\ud{\rho^2}}\rho h(\rho)>0$) translate to the following assumptions in Lagrangian variables: $\frac{\ud{\hat{U}}}{\ud{\vv}}>0$, $\frac{\ud{}^2\hat{U}}{\ud{\vv^2}}<0$, $\frac{\ud{\hat{h}}}{\ud{\vv}}<0$, and $\frac{\ud{}^2\hat{h}}{\ud{\vv^2}}>0$.
For simplicity, we now omit the hats, 
unless explicitly required for clarity. The characteristic speeds of \eqref{eq:ARZ_model_lagrangian} are $\lambda_1 = h'(\vv)$ and $\lambda_2 = 0$, and the associated Rankine-Hugoniot shock jump conditions are
\begin{equation}
\label{eq:ARZ_model_lagrangian_rankine_hugoniot}
\begin{split}
m\brk{\vv}-\brk{u} &= 0\;,\\
\brk{u}+\brk{h(\vv)} &= 0\;,
\end{split}
\end{equation}
where $-m$ is the propagation speed of the shock in the Lagrangian variables (in the Eulerian frame $m$ is the flux of vehicles through the shock). Note that, for contact discontinuities, the conditions are: $m=0$ and $\brk{u} = 0$.

Below, we are going to employ both types of (equivalent) descriptions of the ARZ model. Eulerian \eqref{eq:ARZ_model_conservative} for the computational study of the nonlinear model in \S\ref{sec:computational_study}, and Lagrangian \eqref{eq:ARZ_model_lagrangian} for the jamiton stability analysis in \S\ref{sec:jamiton_stability_analysis}.

Jamiton solutions can now be constructed via the Zel'dovich-von~Neumann-D{\"o}ring (ZND) theory \cite{FickettDavis1979}. One starts out with a traveling wave ansatz. In Eulerian variables, one seeks for solutions $\rho(x,t) = \rho(\eta)$, $u(x,t) = u(\eta)$ of \eqref{eq:ARZ_model_conservative} that depend on the single variable $\eta = \frac{x-st}{\tau}$. In Lagrangian variables, one considers solutions $\vv(\sigma,t) = \vv(\chi)$, $u(\sigma,t) = u(\chi)$ of \eqref{eq:ARZ_model_lagrangian}, where $\chi = \frac{\sigma+mt}{\tau}$. Here $s$ is the traveling wave speed in the road frame, while the Lagrangian wave speed $-m$ relates to the mass flux $m$ of vehicles through the wave.

We start with the Lagrangian formulation \cite{SeiboldFlynnKasimovRosales2013}. The traveling wave ansatz leads to
\begin{align}
\label{eq:ARZ_lagrangian_traveling_wave_1}
\tfrac{m}{\tau}\vv'(\chi) - \tfrac{1}{\tau}u'(\chi) &= 0\;, \\
\label{eq:ARZ_lagrangian_traveling_wave_2}
\tfrac{m}{\tau}u'(\chi) + h'(\vv(\chi))\tfrac{m}{\tau}\vv'(\chi) &= \tfrac{1}{\tau}\prn{U(\vv(\chi))-u(\chi)}\;,
\end{align}
Equation \eqref{eq:ARZ_lagrangian_traveling_wave_1} yields that
\begin{equation}
\label{eq:relation_v_u}
m\vv-u = -s\;,
\end{equation}
where $s$ is a constant of integration. Using \eqref{eq:relation_v_u} to substitute $u$ by $\vv$ in \eqref{eq:ARZ_lagrangian_traveling_wave_2}, we obtain the scalar first-order \emph{jamiton ODE}
\begin{equation}
\label{eq:jamiton_ode}
\vv'(\chi) = \frac{w(\vv(\chi))}{r'(\vv(\chi))}\;,
\end{equation}
where the two functions $w$ and $r$ are defined as
\begin{equation*}
w(\vv) = U(\vv)-(m\vv+s)
\quad\text{and}\quad
r(\vv) = mh(\vv)+m^2\vv\;.
\end{equation*}
Because $h'(\vv)<0$ and $h''(\vv)>0$, the denominator in \eqref{eq:jamiton_ode} has exactly one root, the \emph{sonic value} $\vS$ (occurring at the \emph{sonic point}), such that $h'(\vS) = -m$. The ODE \eqref{eq:jamiton_ode} can be integrated through $\vS$ if the numerator in \eqref{eq:jamiton_ode} has a simple root at $\vS$ as well. This leads to the \emph{Chapman-Jouguet condition} \cite{FickettDavis1979}
\begin{equation*}
m\vS+s = U(\vS)\;,
\end{equation*}
which yields a relationship between the constants $m$ and $s$ as follows:
\begin{equation*}
m = -h'(\vS) \quad\text{and}\quad s = U(\vS)-m\vS\;.
\end{equation*}
One therefore has a one-parameter family of smooth traveling wave solutions, parameterized by $\vS$, each being solutions of \eqref{eq:jamiton_ode}.

Into these smooth profiles shocks can be inserted that move with the same speed $-m$. The first condition in \eqref{eq:ARZ_model_lagrangian_rankine_hugoniot} implies that the quantity $m\vv-u$ is conserved across the shock (in addition to being conserved along the smooth parts by \eqref{eq:relation_v_u}). And both conditions in \eqref{eq:ARZ_model_lagrangian_rankine_hugoniot} together imply that $r(\vv)$ is conserved across shocks. Hence, when integrating \eqref{eq:jamiton_ode}, one can at any value $\vv^-$ insert a shock that jumps to a value $\vv^+$ with $r(\vv^+) = r(\vv^-)$ and continue integrating \eqref{eq:jamiton_ode} from there. Moreover, for those shocks to satisfy the Lax entropy conditions \cite{Whitham1974}, one can only jump downwards, i.e., $\vv^+ < \vS < \vv^-$. This, in turn requires that the smooth jamiton profile $\vv(\chi)$ must be an increasing function. Using L'H{\^o}pital's rule in \eqref{eq:jamiton_ode} at the sonic point yields that
\begin{equation*}
0 < \frac{w'(\vS)}{r''(\vS)} = \frac{U'(\vS)-m}{mh''(\vS)} = \frac{U'(\vS)+h'(\vS)}{mh''(\vS)}\;,
\end{equation*}
which means exactly that the SCC is violated. In other words, as shown in \cite{SeiboldFlynnKasimovRosales2013}, jamiton profiles with shocks can exist if and only if the SCC is violated.

The construction in Eulerian variables is analogous, albeit a bit more technical (cf.~\cite{FlynnKasimovNaveRosalesSeibold2009}). The traveling wave ansatz leads to
\begin{align*}
-s\rho' + (\rho u)' &= 0\;, \\
(u-s-\rho h'(\rho))u' &= U(\rho)-u\;.
\end{align*}
Integrating the first equation yields $\rho(u-s) = m$, which allows one to substitute $\rho$ via $u$ and vice versa. The second equation becomes the jamiton ODE
\begin{equation*}
u'(\eta) = \frac{(u-s)(U(\rho)-u)}{(u-s)^2-m h'(\rho)}\;,
\end{equation*}
where $\rho = \frac{m}{u-s}$. The Chapman-Jouguet condition (matching roots of numerator and denominator) leads to the relations: $m = \rhoS^2 h'(\rhoS)$ and $s = U(\rhoS)-\rhoS h'(\rhoS)$. Shock and entropy conditions are then implemented analogous to the Langrangian situation.

With these rules, jamiton solutions can be constructed (in either choice of variables). For a given choice of $\vS$ (and thus uniform propagation speed), any pattern of solutions to \eqref{eq:jamiton_ode} connected by shocks (satisfying the above conditions) results in a traveling wave solution. The jamitons between any two shocks can be arbitrarily short (with a small variation around $\vS$), or may be arbitrarily long. In fact, it is not even required for the jamitons between shocks to have the same length (see
\cite{FlynnKasimovNaveRosalesSeibold2009,SeiboldFlynnKasimovRosales2013}
for visualizations of jamiton profiles).

While all of these constitute feasible traveling wave solutions of the ARZ model \eqref{eq:ARZ_model_conservative}, it does not mean that all such profiles would be dynamically stable under perturbations. In fact, both numerical evidence (see \S\ref{sec:computational_study}) as well as intuition dictate that neither very short, nor very long jamitons should be stable. The former because they can be thought of as a small (sawtooth) perturbation of the constant $\vS$ state (which is unstable because the SCC is violated, see \S\ref{subsec:stability_uniform_flow}); and the latter because their long tail will itself be close to a constant which, if that state violates the SCC, will be dynamically unstable. In other words, too short jamitons merge and have longer waves form between them; and long jamitons have new instabilities grow in their tails. It is only the middle range of jamitons (not too short and not too long) that is expected to be dynamically stable; and only those should arise in actual practice.

This dynamic stability (of the jamitons themselves) has not been studied before. We do so, by first conducting a computational study in \S\ref{sec:computational_study} that confirms the intuition above and quantifies it; and then deriving and analyzing linear perturbation equations for the jamiton solutions in \S\ref{sec:jamiton_stability_analysis}.

\section{Computational Study of Jamiton Stability}
\label{sec:computational_study}
To understand the dynamic stability of jamitons, we conduct a systematic study of the ARZ model \eqref{eq:ARZ_model_conservative} via direct numerical computation. After constructing a periodic jamiton as outlined in \S\ref{subsec:jamiton_construction}, we insert that profile as an initial condition into a numerical scheme (\S\ref{subsec:computational_study_scheme}) and investigate whether the profile is maintained under small perturbations (\S\ref{subsec:computational_study_results}).

\subsection{Numerical scheme for the ARZ model with relaxation term}
\label{subsec:computational_study_scheme}
The ARZ model \eqref{eq:ARZ_model_conservative} is a system of hyperbolic conservation laws with a relaxation term. The hyperbolic part of the system can be solved using a finite volume scheme based on an approximate Riemann solver \cite{LeVeque2002}. To find the numerical flux at the cell boundaries, we use the HLL approximate Riemann solver \cite{Harten1983}, which guarantees that the numerical fluxes satisfy the entropy condition \cite{lax1973hyperbolic}. Given the grid cell $C_i = [x_i-\Delta x/2,x_i+\Delta x/2]$, where $\Delta x$ is the cell size, let
\begin{equation*}
U_i^n = \begin{bmatrix}
\rho_i^n\\[0.4em] q_i^n
\end{bmatrix}
\quad\text{and}\quad
F^n_{i+\frac{1}{2}}= \begin{bmatrix}
(F_\rho)^n_{i+\frac{1}{2}}\\ (F_q)^n_{i+\frac{1}{2}}
\end{bmatrix}
\end{equation*}
denote the approximate solution (cell average) in cell $C_i$ and the numerical flux at the boundary between cells $C_i$ and $C_{i+1}$, respectively, at time $n\Delta t$ ($n$-th time step).

A numerically robust treatment of the relaxation term is achieved by treating it implicitly, resulting in the semi-implicit update rule
\begin{equation*}
\begin{bmatrix}
\rho_i^{n+1}\\[0.4em] q_i^{n+1}
\end{bmatrix} =
\begin{bmatrix}
\rho_i^{n}\\[0.4em] q_i^{n}
\end{bmatrix}-\tfrac{\Delta t}{\Delta x}\prn{
\begin{bmatrix}
(F_\rho)^n_{i+\frac{1}{2}}\\ (F_q)^n_{i+\frac{1}{2}}
\end{bmatrix} -
\begin{bmatrix}
(F_\rho)^n_{i-\frac{1}{2}}\\ (F_q)^n_{i-\frac{1}{2}}
\end{bmatrix}}
+ \tfrac{\Delta t}{\tau}\!
\begin{bmatrix}
0\\ \rho_i^{n+1} \Big(U(\rho_i^{n+1})+h(\rho_i^{n+1})\Big) -q_i^{n+1}
\end{bmatrix}.
\end{equation*}
This ensures stability even when $\tau$ is small. Note that, because the implicit term appears only in the $q$-equation and because it is linear in $q_i^{n+1}$, the formally semi-implicit numerical scheme is actually fully explicit and the update step can be conducted in two sub-steps:
\begin{enumerate}[ 1)]
\item Update the $\rho$ component explicitly:
\begin{equation*}
\rho_i^{n+1} = \rho_i^n -\tfrac{\Delta t}{\Delta x} \prn{ (F_\rho)^n_{i+\frac{1}{2}} - (F_\rho)^n_{i-\frac{1}{2}} }.
\end{equation*}
\item Now, with $\rho_i^{n+1}$ known from the first step, update
\begin{equation*}
\prn{1-\tfrac{\Delta t}{\tau} } q_i^{n+1}
= q_i^n -\tfrac{\Delta t}{\Delta x}
\prn{(F_q)^n_{i+\frac{1}{2}} - (F_q)^n_{i-\frac{1}{2}}}
+ \tfrac{\Delta t}{\tau} \rho_i^{n+1} \prn{U(\rho_i^{n+1})+h(\rho_i^{n+1})}.
\end{equation*}
\end{enumerate}

\subsection{Results on the stability of jamitons}
\label{subsec:computational_study_results}
Using the numerical scheme described above, we conduct a computational investigation of the stability of jamitons (of the ARZ model \eqref{eq:ARZ_model_conservative} with the specific model functions and parameters described in \S\ref{subsec:specific_model}). Specifically, we classify the jamitons as follows: Evolve the solution up to some large final time, while regularly adding small perturbations. Then a jamiton is classified as stable if the jamiton profile is (within a tolerance) maintained at the final time, and unstable otherwise.

To classify a given jamiton $J_0 = [\rho_0(x), u_0(x)]^T$ (of length $L_0$, with sonic density $\rho_{s_0}$, upstream density $\rho_0^+$, and speed $s_0$), we set up a periodic domain of length $4L_0$ with initial conditions $[\rho_{\text{ic}}(x),u_{\text{ic}}(x)]^T = [\rho_0(x \mod L_0), u_0(x \mod L_0)]^T$, i.e., the initial profile is four consecutive jamitons $J_0$ with shocks in between. We discretize using 10,000 grid cells, and run the numerical scheme (from \S\ref{subsec:computational_study_scheme}) up to $t_\text{final} = $ 3,000 (seconds; we omit units below).

\begin{figure}
\begin{subfigure}{.5\textwidth}
  \centering\captionsetup{width=.96\linewidth}%
  \includegraphics[width=.99\linewidth]{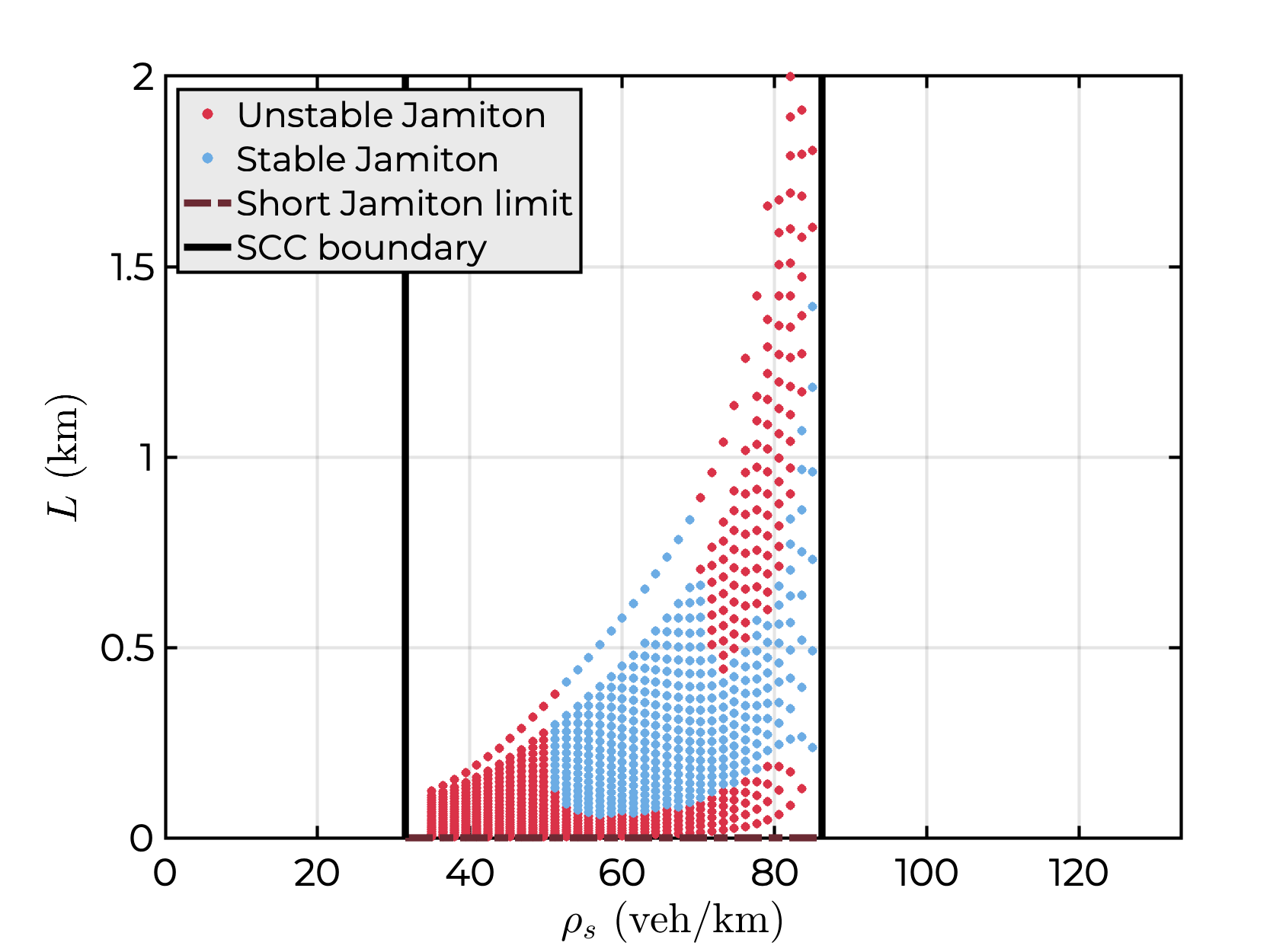}
  \caption{Stability classification in the phase plane $(\rho_\text{S},L)$.}
  \label{subfig:stability_scatter_rhos_vs_L}
\end{subfigure}%
\begin{subfigure}{.5\textwidth}
  \centering\captionsetup{width=.96\linewidth}%
  \includegraphics[width=.99\linewidth]{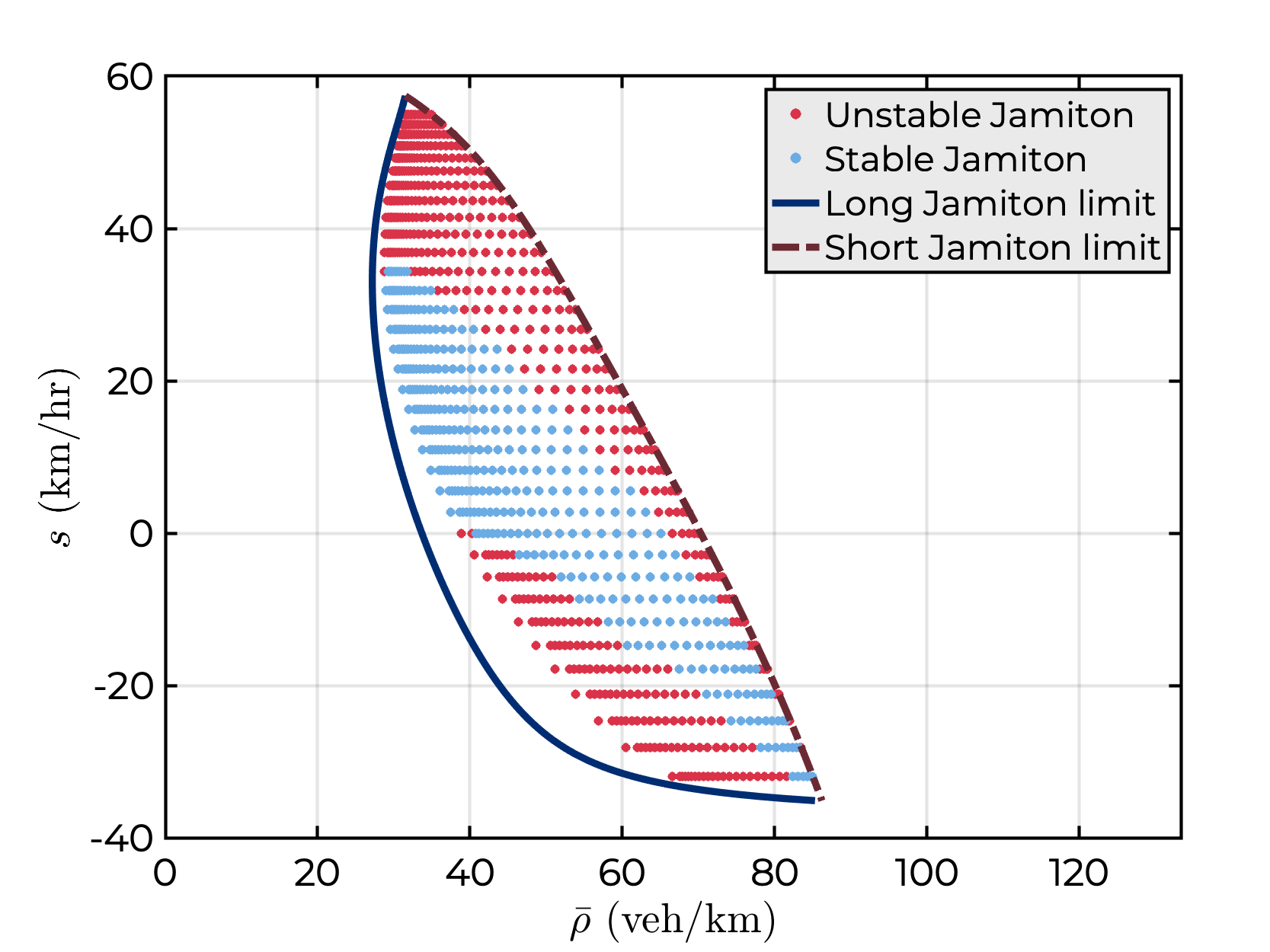}
  \caption{Stability classification in the phase plane $(\bar{\rho},s)$.}
  \label{subfig:stability_scatter_rhobar_vs_s}
\end{subfigure}
\\
\begin{subfigure}{.5\textwidth}
  \centering\captionsetup{width=.96\linewidth}%
  \includegraphics[width=.99\linewidth]{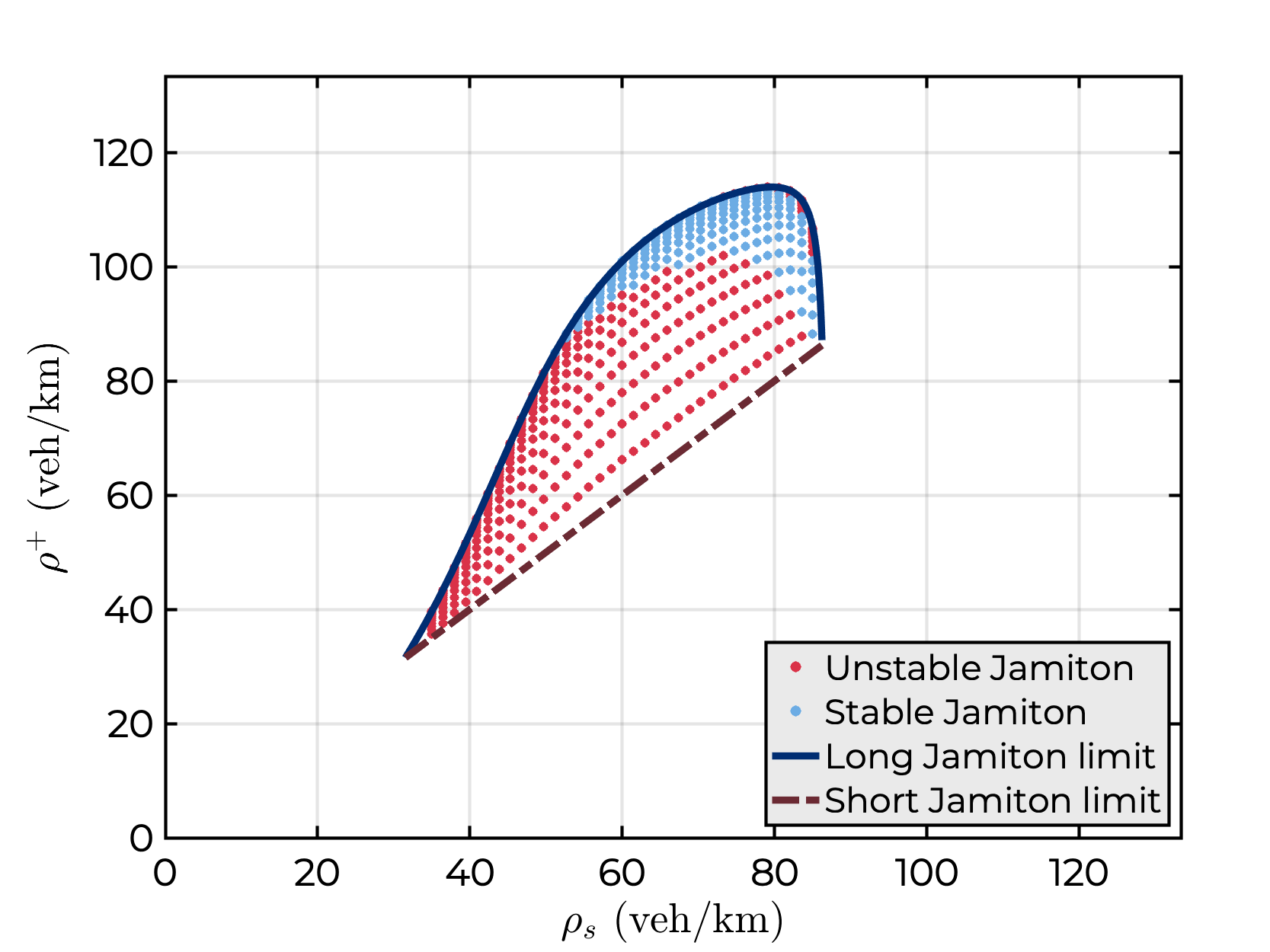}
  \caption{Stability classification in the phase plane $(\rho_\text{S},\rho^+)$.}
  \label{subfig:stability_scatter_rhos_vs_rhop}
\end{subfigure}%
\begin{subfigure}{.5\textwidth}
  \centering\captionsetup{width=.96\linewidth}%
  \includegraphics[width=.99\linewidth]{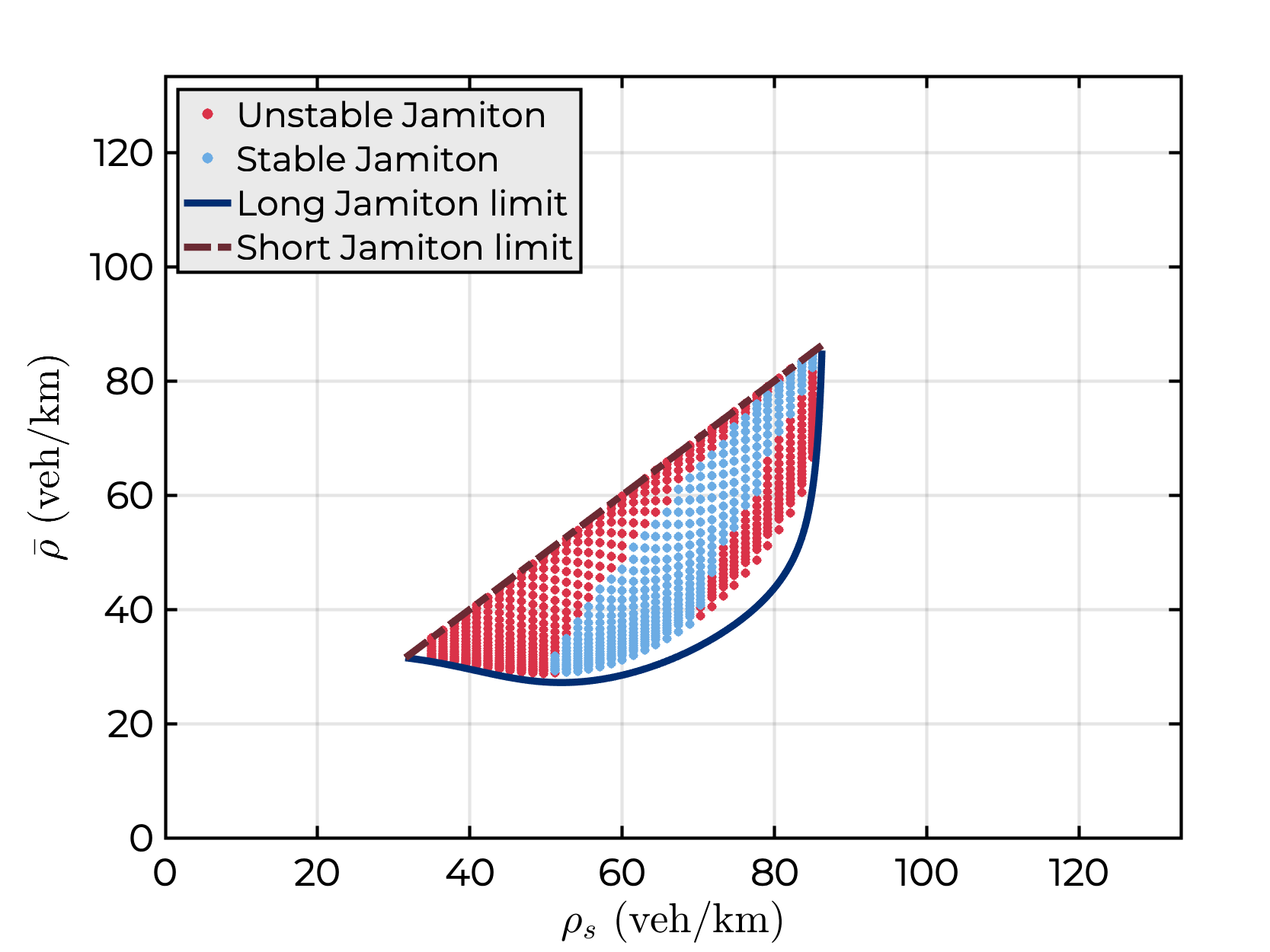}
  \caption{Stability classification in the phase plane $(\rho_\text{S},\bar{\rho})$.}
  \label{subfig:stability_scatter_rhos_vs_rhobar}
\end{subfigure}
\caption{Classification of 980 jamitons into stable and unstable, displayed in four different phase planes. In each plane, the dashed brown line represents the zero length jamiton, and the dark blue line is the limit of jamitons with infinite length. The two disconnected red regions correspond to the ``splitting'' and ``merging'' instabilities, respectively.}
\label{fig:stability_scatter}
\end{figure}

During the numerical solution process, a small smooth perturbation is added to the vehicle velocity field $u = q/\rho - h(\rho)$ in each step. The perturbation in the $n$-th step is
\begin{equation*}
p^n(x) = \sqrt{\Delta t}\, c(t) \frac{1}{\sqrt{\ell}}
\sum_{\nu=1}^\ell \xi^n_\nu \sin \prn{\frac{2 \pi \nu x}{L_0}},
\end{equation*}
where the $\xi^n_\nu\in \mathcal{N}(0,1)$ are normally distributed random numbers with mean zero and standard deviation 1. As in the Euler-Maruyama method, the additive noise is scaled with $\sqrt{\Delta t}$. The value $\ell$ is chosen so that the highest frequency mode has a period $\frac{L_0}{\ell}$ that is not below the vehicle length $1/\rho_\text{max}$, i.e., $\ell = \lfloor L_0\rho_\text{max} \rfloor$. In other words, we have white noise exactly until the vehicle scale, which is well-resolved by the numerical scheme. Finally, the noise scale is $c(t) = \tfrac{1}{100} u_\text{max}$ for $t\le 100$, and $c(t) = \tfrac{1}{1000} u_\text{max}$ for $t>100$. The rationale for this larger initial ``thermal noise'' is, like in probabilistic optimization techniques, to make it easier for the solutions to escape their initial configuration in case it is only mildly unstable.

Once the solution at $t_\text{final}$ is found, we first determine the number of shocks. If that number is not equal to 4, we immediately classify the jamiton $J_0$ as unstable. Otherwise, we check the jamiton speed $s$ by plotting the points $(\rho(x_i,t_\text{final}),\rho(x_i,t_\text{final})u(x_i,t_\text{final}))$ for $i=1,\dots,10000$ in the fundamental diagram (FD), and calculate $s$ as the least squares best fit slope of these data points (see \cite{SeiboldFlynnKasimovRosales2013} for the reason why $s$ is the slope in the FD). If $|s-s_0|>0.5\text{m}/\text{s}$, we classify $J_0$ as unstable. Otherwise, we classify $J_0$ as stable.

This process is now conducted (and run in parallel on a HPC cluster) for 980 different jamitons that are sampled as follows. First we sample 35 values of $\rhoS$ equidistant in the $\rho$-interval where the SCC is violated. Then, for each $\rhoS$, we pick 28 values of $\rho^+$ in $[\rhoS,\rho_\text{M}]$, where $\rho_\text{M}$ is the upstream density corresponding to the infinite jamiton \cite{SeiboldFlynnKasimovRosales2013}.

The results of this classification are displayed in Fig.~\ref{fig:stability_scatter}. Each of the four panels shows the same results, but in four different ``phase planes''. Each jamiton is uniquely determined by two parameters: (i)~the sonic density $\rhoS$ or equivalently the wave speed $s$; and (ii)~the downstream shock density $\rho^+$, or equivalently, the average density $\bar{\rho}$ across the jamiton, or equivalently, the jamiton length $L$. Panels~\ref{subfig:stability_scatter_rhos_vs_L}, \ref{subfig:stability_scatter_rhos_vs_rhop}, and~\ref{subfig:stability_scatter_rhos_vs_rhobar} have the $\rhoS$ on the horizontal axis, and $L$, $\rho^+$, and $\bar{\rho}$, respectively, on the vertical axis. Panel~\ref{subfig:stability_scatter_rhobar_vs_s} displays $s$ vs.~$\bar{\rho}$. In each quantity except $L$, the jamiton region (where the SCC \eqref{eq:linear_stability_condition} is violated) spans an interval. The dashed brown curve corresponds the zero-length jamiton limit (in which $\rhoS = \rho^+ = \bar{\rho}$), while the solid dark blue curve represents the limit of infinitely long jamitons. Inside that jamiton domain, the 980 investigated jamitons are displayed as colored dots: stable jamitons are light blue; unstable jamitons are red. Note that the void regions visible in Panel~\ref{subfig:stability_scatter_rhos_vs_L} (top left), Panel~\ref{subfig:stability_scatter_rhobar_vs_s} (bottom left), and Panel~\ref{subfig:stability_scatter_rhos_vs_rhobar} (bottom right), also possess jamitons that were not simulated due to the sampling strategy of the 980 examples.

The results display intriguingly clear patterns: there appear to be two smooth curves inside the jamiton region that separate the stable from the unstable jamitons. Specifically, there are two unstable regions separated by a stable region: short jamitons which perturbations cause to coalesce into bigger ones (a ``merging'' instability); and long jamitons in which the long tail is linearly unstable and sheds growing waves (a ``splitting'' instability). This last characterization of these two mechanisms is based on observing the time-evolution of the computations, as well as the stability analysis below.

\section{Stability Analysis of Jamiton Solutions}
\label{sec:jamiton_stability_analysis}
We now move towards a mathematical analysis of the dynamic stability of jamitons. For this we switch to the Langrangian variables introduced in \S\ref{subsec:jamiton_construction}. Consider a given jamiton $[\vv_0(\sigma,t)$, $u_0(\sigma,t)]^T$ with sonic specific volume $\vv_{s_0}$, and Lagrangian length (which is actually the number of vehicles in the jamiton) $N_0$. We start by writing the (Lagrangrian) ARZ model \eqref{eq:ARZ_model_lagrangian} in the frame of reference of this jamiton, which has a propagation speed $-m_0 = h'(\vv_{s_0})$. Thus we introduce the variables $\chi = \frac{\sigma + m_0\/t}{\tau}$ (the same variable used in \S\ref{subsec:jamiton_construction} to construct the jamitons) and the non-dimensional time $\tvar = \frac{t}{\tau}$ (for consistency with the scaling used for $\chi$). Because of that last choice, any instability growth rate computed with these variables needs to be scaled by $\tau$ to recover physical units.

In the coordinates defined above, equations \eqref{eq:ARZ_model_lagrangian} become
\begin{equation}
\label{eq:ARZ_lagrangian_variables_FoR}
\begin{split}
\vv_{\tvar}+(m_0 \vv -u)_{\chi} &= 0\;,\\
\prn{u+h(\vv)}_{\tvar} + m_0 \prn{u+h(\vv)}_{\chi} &= U(\vv)-u\;.
\end{split}
\end{equation}
This system is in conservative form, with conserved quantities $\vv$
and $q=u+h(\vv)$. The characteristic speeds of
\eqref{eq:ARZ_lagrangian_variables_FoR} are
\begin{equation}
\label{eq:char_speeds_ARZ_lagrangian_variables_FoR}
\lambda_1 = m_0+h'(\vv)
\quad\text{and}\quad
\lambda_2 = m_0\;.
\end{equation}
The Rankine-Hugoniot shock jump conditions associated with \eqref{eq:ARZ_lagrangian_variables_FoR} are
\begin{equation}
\label{eq:ARZ_lag_FoR_RK_jump_conditions}
\begin{split}
(-m_0+\tilde{m})\brk{\vv} + \brk{u} &=0\;,\\
\brk{u}+\brk{h(\vv)} &= 0\;.
\end{split}
\end{equation}
where $\tilde{m}$ is the shock speed in the $\chi$--$\tvar$ frame. Contacts require $\tilde{m}=m_0$ and $\brk{u}=0$.

\subsection{Perturbation system for single-jamiton waves}
We now formulate a linear perturbation system of \eqref{eq:ARZ_lagrangian_variables_FoR}. There are two fundamental differences to the linear perturbation analysis for uniform flow presented in \S\ref{subsec:stability_uniform_flow}. First, because the jamiton profile is non-constant, we obtain a variable coefficient linear system. Second, because the jamiton contains a shock, we must introduce a perturbation to the shock's position as an additional variable (a variable not needed for perturbations of smooth solutions). As we will see below in more detail, both aspects render this analysis significantly more complicated than the one in \S\ref{subsec:stability_uniform_flow}.

Here we consider the stability of periodic jamiton profiles with one shock per period, under periodic perturbations. Note that this setup excludes the possibility of jamitons merging by means of adjacent shocks approaching each other. Hence, we only study the ``splitting instability'' for long jamitons, not the ``merging instability'' for short jamitons (see \S\ref{subsec:computational_study_results}).

Consider a periodic jamiton profile $[\vv_0(\sigma,t)$, $u_0(\sigma,t)]^T$ of length $N_0$ between shocks, and write it as $[\vv_0(\chi)$, $u_0(\chi)]^T$ --- a solution of \eqref{eq:ARZ_lagrangian_variables_FoR} on $[0,N_0]$ with the shock placed at 0. Now write $\vv(\chi,\tvar) = \vv_0(\chi) + \delta \vv(\chi,\tvar)$ and $u(\chi,\tvar) = u_0(\chi) + \delta u(\chi,\tvar)$, where $\delta \vv$ and $\delta u$ are infinitesimal perturbations. Substituting into \eqref{eq:ARZ_lagrangian_variables_FoR} yields the linear system for $\delta \vv$ and $\delta u$:
\begin{equation}
\label{eq:perturbation_model_1}
\begin{split}
\delta \vv_{\tvar}+(m_0 \delta \vv -\delta u)_{\chi} &= 0\;,\\
\prn{\delta u+h^\pr(\vv_0)\delta \vv}_{\tvar} +
m_0 \prn{\delta u+h^\pr(\vv_0) \delta \vv}_{\chi}
&= U^\pr(\vv_0)\delta \vv - \delta u\;.
\end{split}
\end{equation}
We also need to track the infinitesimal perturbation of the shock position
$\chi = \mu(\tvar)$. We do so by implementing the Rankine-Hugoniot
conditions \eqref{eq:ARZ_lag_FoR_RK_jump_conditions} in a way consistent
with solving \eqref{eq:ARZ_lagrangian_variables_FoR} on $[0,N_0]$ with
periodic boundary conditions. This then generates boundary conditions for \eqref{eq:perturbation_model_1}. The first equation in \eqref{eq:ARZ_lag_FoR_RK_jump_conditions} yields
\begin{equation*}
(\dot{\mu}-m_0)\Big([\vv_0]+[\delta \vv]+\mu[\vv_{0 \chi}]\Big) +[u_0]+[\delta u] +\mu[u_{0 \chi}] = 0\;.
\end{equation*}
Expanding this equation, ignoring terms beyond $O(\mu)$, and using that $[u_0]-m_0[\vv_0] = 0$ and $u_{0\chi} - m_0 \vv_{0\chi} = 0$, we obtain
\begin{equation*}
\dot{\mu}[\vv_0] - m_0 [\delta \vv]+[\delta u] = 0\;.
\end{equation*}
The second equation in \eqref{eq:ARZ_lag_FoR_RK_jump_conditions} becomes
\begin{equation*}
[u_0] + [\delta u] + \mu[u_{0 \chi}] + [\vv_0] + [h'(\vv_0)\delta \vv] + \mu[h(\vv_0)_{\chi}] = 0\;.
\end{equation*}
Again, ignoring terms beyond $O(\mu)$ and using that $[u_0]+[h(\vv_0)] = 0$, we get
\begin{equation*}
[\delta u] + [h'(\vv_0) \delta \vv] + \mu [u_{0 \chi} + h'(\vv_0)_{\chi}] = 0\;.
\end{equation*}
In this setup the bracket notation denotes $\brk{\zeta} = \zeta(0^+)-\zeta(N_0^-)$. Therefore, we have derived the following variable-coefficient linear model for $\delta \vv$ and $\delta u$ on $[0,N_0]$, with boundary conditions that involve the shock position perturbation $\mu$:
\begin{align}
\label{eq:model_dv_du}
\begin{split}
\delta \vv_{\tvar}+(m_0 \delta \vv -\delta u)_{\chi} &= 0\;,\\
\big(\delta u+h'(\vv_0)\delta \vv \big)_{\tvar} + m_0 \big( \delta u+h'(\vv_0) \delta \vv \big)_{\chi}
&= U'(\vv_0)\delta \vv- \delta u\;,
\end{split}\\[.5em]
\nonumber
\text{with boundary condition:}\quad
\brk{\delta u + h'(\vv_0) \delta \vv} &= - \mu \brk{u_{0 \chi} +h'( \vv_0)_{\chi}},\\[.2em]
\label{eq:dv_du_bc}
\text{where } \mu \text{ satisfies the ODE:}\quad\quad
\dot{\mu} &= \frac{m_0 [\delta \vv]-[\delta u]}{[\vv_0]}\;.
\end{align}

We conduct two further simplifications to the model. First, we transform it to characteristic form by writing it in terms of the Riemann variables $\delta u$ and $\delta q = \delta u + h'(\vv_0)\delta \vv$. Second, we replace the shock perturbation variable $\mu$ by a Robin b.c.~for the PDE, as follows. Differentiating the boundary conditions $[\delta q] = - \mu [u_{0 \chi} + h'( \vv_0)_{\chi}]$ with respect to time yields
\begin{equation*}
\frac{d}{d\tvar} \brk{\delta q}
= - \brk{u_{0 \chi} +h'( \vv_0)_{\chi}} \dot{\mu}
= \frac{- \brk{u_{0 \chi} + h'( \vv_0)_{\chi}}}{[\vv_0]}
\prn{\brk{\frac{m_0}{h'(\vv_0)} \delta q}
- \brk{ \prn{1+\frac{m_0}{h'(\vv_0)}} \delta u } }.
\end{equation*}
Using the fact that $\delta q_{\tvar} = -m_0 \delta q_{\chi} +
\prn{\frac{-h'(\vv_0) - U'(\vv_0)}{h'(\vv_0)}}\delta u
+ \prn{\frac{U'(\vv_0)}{h'(\vv_0)}}\delta q$,
we obtain Robin boundary conditions for the PDE. Altogether, we obtain the following system
\begin{equation}
\label{eq:perturbation_system_PDE}
\begin{split}
\delta u_{\tvar} + \prn{m_0+h'(\vv_0)} \delta u_{\chi}
&= \prn{\tfrac{m_0 h'(\vv_0)_\chi-h'(\vv_0) - U'(\vv_0)}{h'(\vv_0)}}\delta u
+ \prn{\tfrac{U'(\vv_0) - m_0 h'(\vv_0)_{\chi}}{h'(\vv_0)} }\delta q\;,\\
\delta q_{\tvar} + m_0 \delta q_{\chi}
&= \prn{\tfrac{-h'(\vv_0) - U'(\vv_0)}{h'(\vv_0)} }\delta u
+ \prn{\tfrac{U'(\vv_0)}{h'(\vv_0)} }\delta q\;,
\end{split}
\end{equation}
with boundary condition
\begin{equation}
\label{eq:perturbation_system_bc}
\delta q_\chi(0) + k_\text{L} \delta q(0) = \delta q_\chi(N_0) + k_\text{R} \delta q(N_0) + c_\text{L} \delta u(0) + c_\text{R} \delta u (N_0)\;.
\end{equation}
The coefficients are computable from the jamiton functions as
\begin{equation*}
k_\text{L} = K(0)\;, \quad
k_\text{R} = K(N_0)\;, \quad
c_\text{L} = -C(0)\;,
\quad\text{and}\quad
c_\text{R} = C(N_0)\;,
\end{equation*}
where
\begin{align*}
K(\chi) &= \tfrac{1}{m_0}
- \tfrac{ \brk{u_{0 \chi} +h'( \text{v}_0)_{\chi}}}{[\text{v}_0]}
\tfrac{1}{h'(\text{v}_0(\chi))}
- \tfrac{h'(\text{v}_0(\chi))+ U'(\text{v}_0(\chi))} {m_0 h'(\text{v}_0(\chi))}\;,
\\
C(\chi) &=
\tfrac{ \brk{u_{0 \chi} +h'( \text{v}_0)_{\chi}}}{[\text{v}_0]}
\prn{\tfrac{1}{m_0}+\tfrac{1}{h'(\text{v}_0(\chi))}}
+ \tfrac{h'(\text{v}_0(\chi))+ U'(\text{v}_0(\chi))} {m_0 h'(\text{v}_0(\chi))}\;.
\end{align*}

\subsection{Qualitative characterization of the jamiton perturbation system}
\begin{figure}
\begin{subfigure}{.33\textwidth}
  \centering
  \includegraphics[width=.999\linewidth]{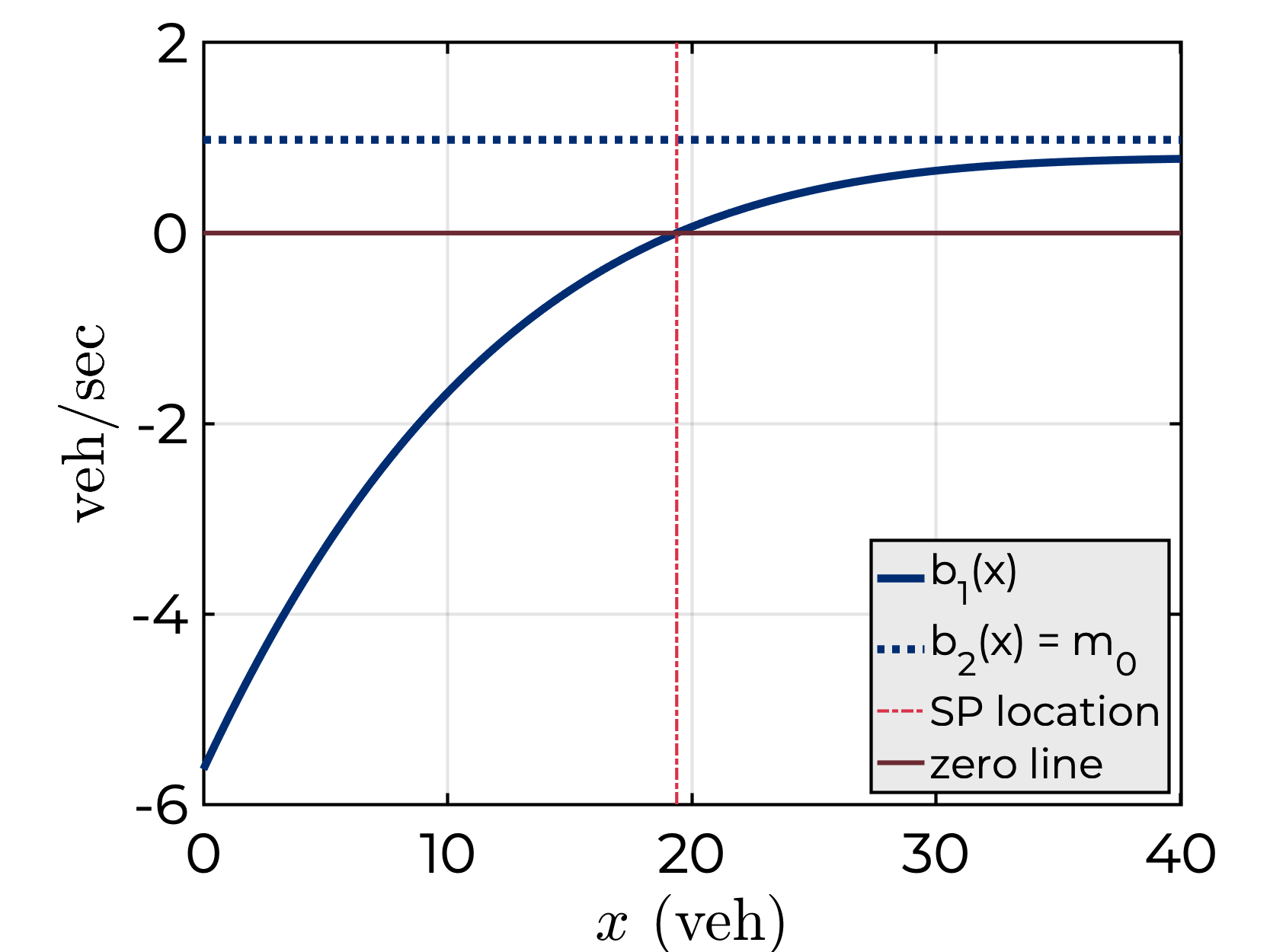}
  \caption{$b_1(x)$ and $b_2$}
  \label{subfig:b}
\end{subfigure}%
\begin{subfigure}{.33\textwidth}
  \centering
  \includegraphics[width=.999\linewidth]{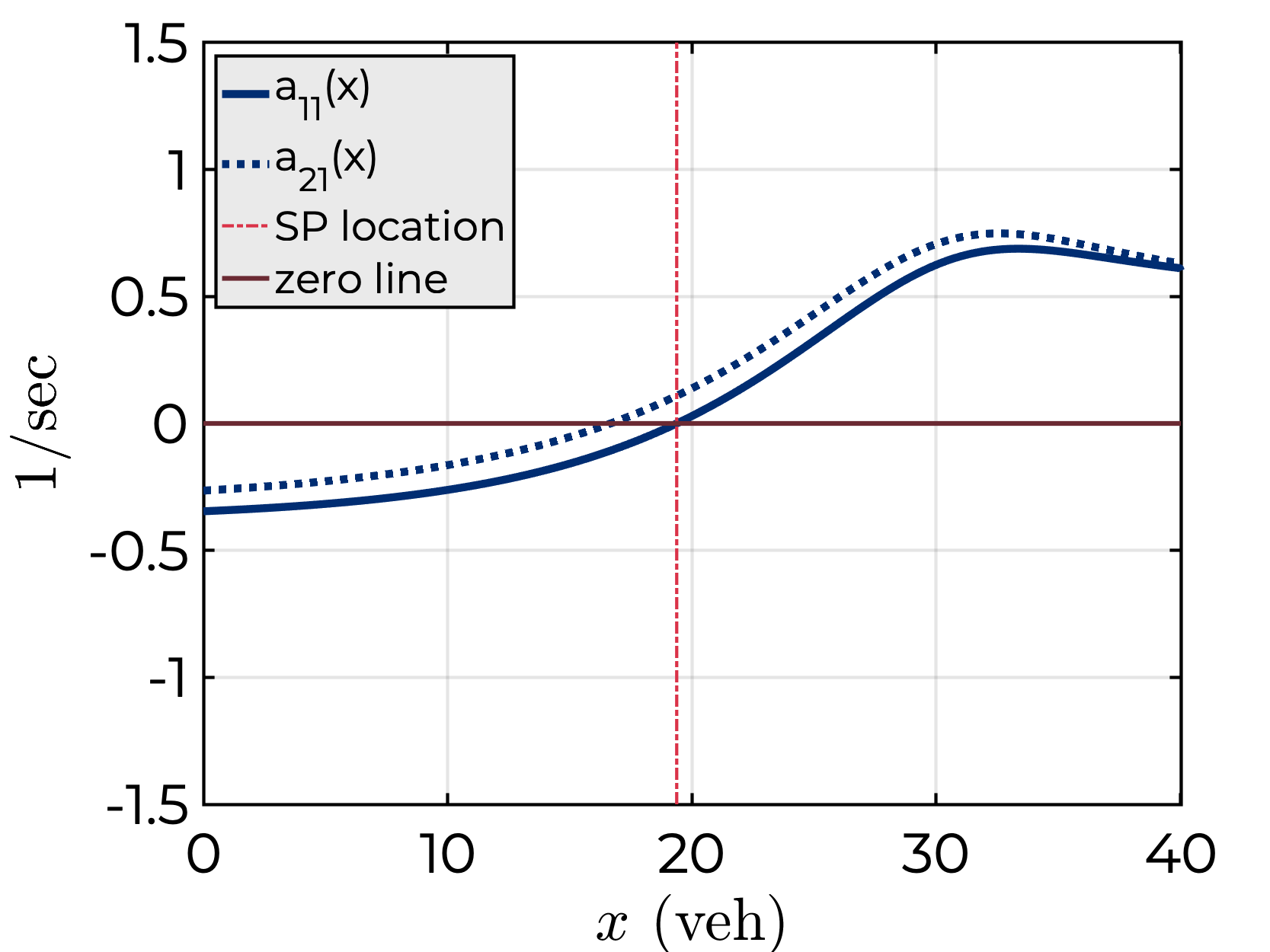}
  \caption{$a_{11}(x)$ and $a_{21}(x)$}
  \label{subfig:a1}
\end{subfigure}%
\begin{subfigure}{.33\textwidth}
  \centering
  \includegraphics[width=.999\linewidth]{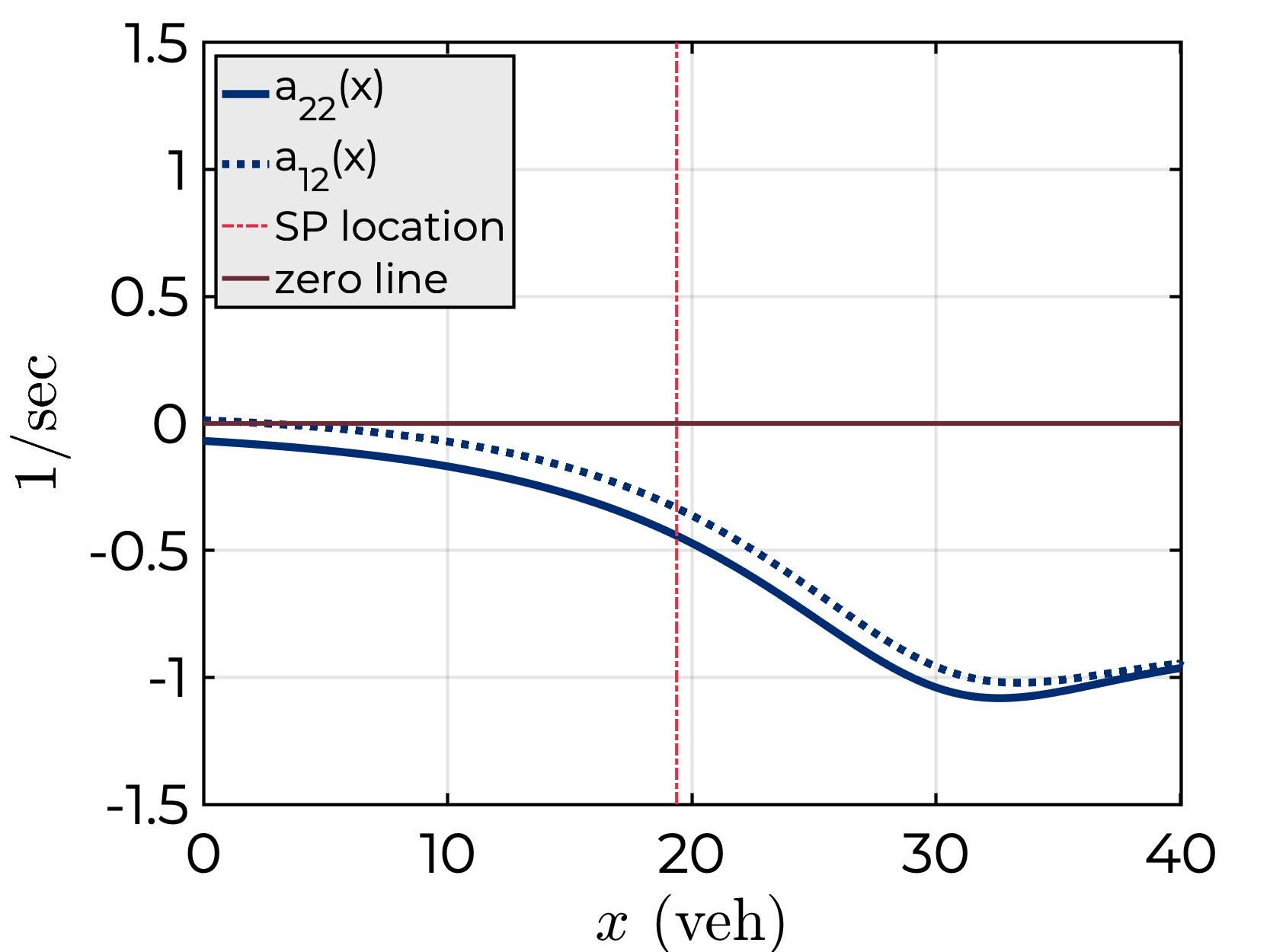}
  \caption{$a_{22}(x)$ and $a_{12}(x)$}
  \label{subfig:a2}
\end{subfigure}
\caption{Coefficient functions for \eqref{eq:perturbation_system_PDE_short}
and a jamiton with $\vS = 12.5$m/veh and $\vv^+ = 8.9$m/veh. This jamiton
has a length of 561m and contains 40 vehicles. Note that here we revert
to physical units (vehicles) for the horizontal axis.}
\label{fig:coefficient_functions}
\end{figure}

We now adopt a short notation for the jamiton perturbation system \eqref{eq:perturbation_system_PDE}, with b.c.~\eqref{eq:perturbation_system_bc}, by writing $(u,q)$ and $(x,t)$ in place of of $(\delta u,\delta q)$ and $(\chi,\tvar)$, and introducing coefficient functions to obtain:
\begin{equation}
\label{eq:perturbation_system_PDE_short}
\begin{split}
u_t + b_1(x) u_x &= a_{11}(x)u + a_{12}(x)q\;,\\
q_t + b_2\phantom{(x)} q_x &= a_{21}(x)u + a_{22}(x)q\;,
\end{split}
\end{equation}
with b.c.~$(q_x + k_\text{L} q)(0) = (q_x + k_\text{R} q)(N_0) + c_\text{L} u(0) + c_\text{R} u (N_0)$. The characteristic speed $b_2 > 0$ is constant and positive. In turn, $b_1(x)$ vanishes at the sonic point $\xS$, and is negative (positive) for $x<\xS$ ($x>\xS$). Hence, the only in-going characteristic is at $x=0$, for $q$ (consistent with a single b.c.). The function $a_{11}(x)$ crosses from negative to positive at $\xS$ as well, and it is always negative for $x<\xS$; it may or may not cross back to negative for some $x>\xS$. Finally, $a_{22}(x)<0$ everywhere. Figures~\ref{fig:coefficient_functions} and~\ref{fig:characteristics} display the functions and characteristic curves, respectively, for an example jamiton.

Qualitatively, the solutions of \eqref{eq:perturbation_system_PDE_short} behave as follows. Being an advection-reaction system, its solutions are generally wave-like in nature. Waves enter the $q$-field at $x=0$ and are transported with the $q$-field to the right with constant speed $b_2$, while being dampened by the $a_{22}$-term and modified (via the $u$-field) through the $a_{21}$-term. Likewise, the $q$-field constantly feeds into the $u$-field via the $a_{12}$-term. Moreover, for $x<\xS$, the $u$-field is transported towards $x=0$ and dampened by $a_{11}$; while for $x>\xS$, the $u$-field is transported towards $x=N_0$ and amplified/dampened by $a_{11}$. Finally, the outgoing characteristics at $x=0$ ($u$) and $x=N_0$ ($u$ and $q$) combine via \eqref{eq:perturbation_system_bc} and feed back into $q$ at $x=0$.

Our goal is now to (a)~characterize the dynamic stability of the given jamiton by means of the behavior of the solutions of it associated perturbation system \eqref{eq:perturbation_system_PDE_short} (incl.~b.c.), and (b)~use this insight to explain and understand the computational results of the fully nonlinear ARZ model \eqref{eq:ARZ_model_conservative} presented in \S\ref{sec:computational_study}. To that end, we start by establishing that there are (at least) two distinct notions of (in)stability that must be considered here.

First, \emph{asymptotic stability} under infinitesimal perturbations (studied in \S\ref{subsec:asymptotic_stability}). This is captured by the $t\to\infty$ behavior of linear model \eqref{eq:perturbation_system_PDE_short}: if for any i.c.~$[u(x,0),q(x,0)]^T$ the solution decays exponentially as $t\to\infty$, then this notion of stability is met. \emph{Strong linear instability} occurs when there is a positive feedback mechanism that produces an exponential growth of an initial perturbation in time, eventually driving the full model \eqref{eq:ARZ_model_conservative} out of the linear regime, no matter how small the initial (non-zero) perturbation is. At the borderline between these two behaviors, the solutions to the linear system may remain bounded for all time, or grow/decay at a sub-exponential rate.

The second notion of stability is given by the \emph{maximum transient growth} criteria (studied in \S\ref{subsec:transient_growth}). Because \eqref{eq:perturbation_system_PDE_short} is non-normal, even if asymptotic stability applies, an initially small perturbation may be amplified significantly at transient times, before eventually dying off as $t\to\infty$. However, if that amplified perturbation becomes sufficiently large, nonlinear effects will take over in the full ARZ model \eqref{eq:ARZ_model_conservative}. In this scenario, how far the system ends up from equilibrium depends both on the transient growth factor (see below) and the magnitude of the perturbations.

\subsection{Fundamental challenges caused by the sonic point}
\label{subsec:fundamental_challenges}
In the same way as the original inhomogenous ARZ model may look misleadingly innocuous (``just a hyperbolic system with a relaxation term''), yet develops extremely complex dynamics if the SCC is violated, the jamiton perturbation system \eqref{eq:perturbation_system_PDE} may look innocent as well --- and also that impression would be false. The fact that the characteristic speed $b_1$ transitions from negative to positive at $\xS$ (a direct consequence of this being a sonic point), causes fundamental structural challenges.

It may seem rather natural to attempt to study \eqref{eq:perturbation_system_PDE} by expanding its solutions using eigenmodes, and seek solutions to the eigenvalue problem
\begin{equation}
\label{eq:eigenvalue_problem}
\begin{cases}
\lambda u &= -b_1(x) u_x + a_{11}(x)u + a_{12}(x)q\;,\\
\lambda q &= -b_2 q_x + a_{21}(x)u + a_{22}(x)q\;.
\end{cases}
\end{equation}
However, the right hand side operator here is non-normal; and it is well known that for non-normal operators, spectral calculations can be extremely unreliable \cite{TrefethenSept1997, TrefethenJan1999, EmbreeTrefethen2001}.

Furthermore, the presence of the sonic point makes the situation substantially worse, even if one were to have access to ``exact'' computations. To illustrate the issue consider the simple model problem
\begin{equation}
\label{eq:simple_model_problem}
u_t + (x u)_x = \tfrac{3}{4} u\;, \quad -1 < x < 1\;.
\end{equation}
The exact solution of \eqref{eq:simple_model_problem} is easily obtained using characteristics: $u = u_0(x\,e^{-t})e^{-\frac{1}{4}t}$, where $u_0$ is the initial data. This \emph{clearly} is a stable situation by any ``physically reasonable'' definition. On the other hand, if we look for eigenfunctions by separating $u = \phi(x)\/e^{\lambda\,t}$, we find that: $\phi = |x|^\alpha$, with $\alpha = -(\lambda+\frac{1}{4})$ and \emph{any} $\lambda$ with $\text{Re}(\lambda) < \frac{1}{4}$, is an acceptable square-integrable eigenfunction. Even worse: every eigenvalue has infinite multiplicity (apply $\frac{\ud{}^n}{\ud{\alpha^n}}$ to the eigenvalue equation with the solutions above).

Thus from a naive eigenvalue calculation one would conclude that an exponential instability occurs! But here, with an exact solution, the situation is clear: the presence of a sonic point allows the existence of solutions that are not smooth. Then stability and growth/decay rates depend on the smoothness restrictions imposed. While $L^2$ yields instability, $L^\infty$ or $H^1$ yield stability, but with different bounds on the decay rates. Thus, in a numerical computations one would have to worry about what restriction (if any) the computation enforces as the resolution increases.

Because of these issues we refrain from using the approach in \eqref{eq:eigenvalue_problem}, and instead characterize (in)stability via alternative ways that do not use eigenmode expansions.

\subsection{Quantitative results: Asymptotic stability}
\label{subsec:asymptotic_stability}
The $t\to\infty$ behavior of the solutions of the jamiton perturbation system \eqref{eq:perturbation_system_PDE_short} (incl.~b.c.) depends on a delicate balance of growth vs.\ decay effects. And because those are governed by the functions $a_{ij}(x)$, $b_i(x)$, and the b.c.~constants, we do not attempt a fully analytical characterization here. Instead, we formulate a sequence of approximations to the solutions of \eqref{eq:perturbation_system_PDE_short} and analyze their behavior. Specifically, we formulate the following approximation scheme.

\begin{figure}
\begin{subfigure}{.5\textwidth}
  \centering\captionsetup{width=.85\linewidth}%
  \includegraphics[width=.99\linewidth]{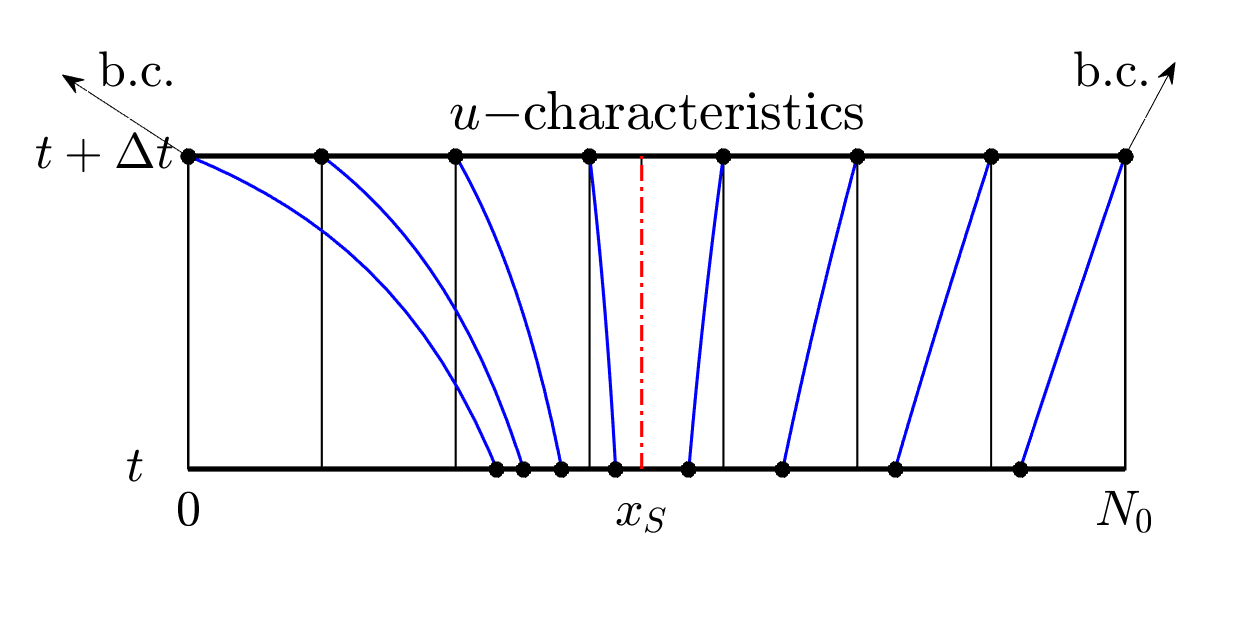}
  \label{subfig:u-characteristics}
\end{subfigure}%
\begin{subfigure}{.5\textwidth}
  \centering\captionsetup{width=.85\linewidth}%
  \includegraphics[width=.99\linewidth]{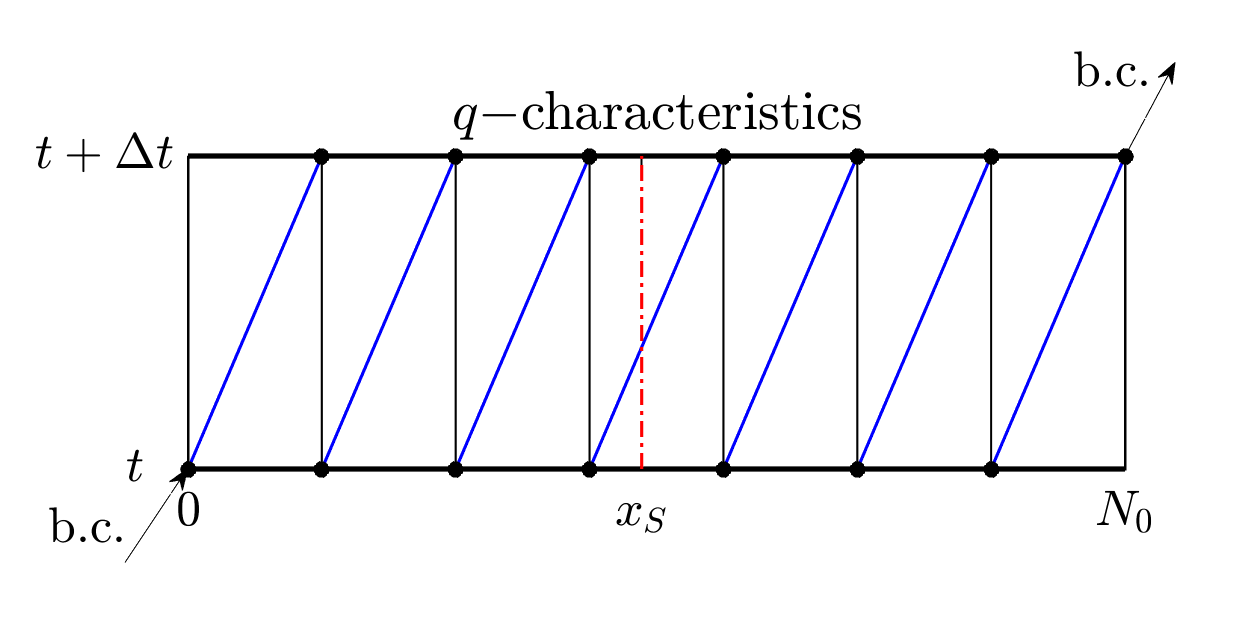}
  \label{subfig:q-characteristics}
\end{subfigure}
\vspace{-2.5em}
\caption{Illustration of the discretization used to approximate \eqref{eq:perturbation_system_PDE_short}, as described in \S\ref{subsec:asymptotic_stability}. The left (right) graphic shows the characteristic curves corresponding to the $u$ ($q$) variable. The $u$-characteristics expand away from the sonic point towards the domain boundaries (where the shock is). The scheme's time step is selected so that the $q$-characteristics advance by $h$ per time step.}
\label{fig:characteristics}
\end{figure}

We discretize the spatial domain into a regular grid $\{x_0,\dots,x_m\} = \{0,h,2h,\dots,N_0-h,N_0\}$ and conduct time steps of size $\Delta t = h/b_2$, see Fig.~\ref{fig:characteristics}. We denote the grid approximations $U_j^n \approx u(jh,n\Delta t)$ and $Q_j^n \approx q(jh,n\Delta t)$, and denote the full state vector at time $n\Delta t$ by $\vec{Y}^n = [\vec{U}^n,\vec{Q}^n]^T$, where $\vec{U}^n = [U_1^n,\dots,U_m^n]^T$ and $\vec{Q}^n = [Q_1^n,\dots,Q_m^n]^T$. An update matrix for the transport part of \eqref{eq:perturbation_system_PDE_short} (incl.~b.c., but neglecting the $a_{ij}$-terms) is obtained via tracking characteristics: for each grid point $x_j = jh$, determine the associated foot point $\mathring{x}_j$ as the solution of the ODE $\dot{x}(s) = -b_1(x(s))$ with $x(0) = x_j$, evaluated at $s = \Delta t$. Then, $U_j^{n+1} = \Psi_{\vec{U}^n}(\mathring{x}_j)$, where $\Psi_{\vec{U}^n}(x)$ is the piecewise-linear interpolant based on the grid data $\vec{U}^n$. Due to the clever choice of time step, the $q$-update can be solved exactly via $Q_j^{n+1} = Q_{j-1}^n$ for all $j\ge 1$. The b.c.~are used to update $Q_0^{n+1} = \frac{1}{k_\text{L} - h^{-1}} ( (k_\text{R}-h^{-1})Q_{m-1}^n +c_\text{L} \Psi_{\vec{U}^n}(\mathring{x}_0) + c_\text{R} \Psi_{\vec{U}^n}(\mathring{x}_m))$. We denote this update matrix $M_1$.

A second matrix for the growth/decay part (i.e., neglecting the advection terms) is formulated as follows:
$[U_j^{n+1},Q_j^{n+1}]^T = \exp(\Delta t A(x_j))\cdot [U_j^n,Q_j^n]^T$, where $A(x)$ is the $2\times 2$ matrix formed by the $a_{ij}(x)$ values. We denote the resulting update matrix $M_2$.

One step of the numerical scheme, $\vec{Y}^{n+1} = M\cdot\vec{Y}^n$, is given by the update matrix $M = M_2\cdot M_1$. This first order method is carefully designed to not incur any slow drifts. Because the scheme is linear with time-independent coefficients, the $t\to\infty$ behavior of the solutions is fully characterized by its one-step update matrix $M$, specifically by its spectral radius $\rho(M)$: asymptotic stability (of the approximation) is given exactly if $\rho(M) < 1$. Once $M$ is set up, this stability condition can be checked via Matlab's numerical linear algebra routines, resulting in a systematic classification of jamitons into asymptotically stable vs.~unstable.

A caveat in this approach is that for any choice of grid size $h$, we check the asymptotic stability of an \emph{approximation} to \eqref{eq:perturbation_system_PDE_short}. However, because we have a convergent sequence of approximations, we approach the true answer for \eqref{eq:perturbation_system_PDE_short} as $h\to 0$. Moreover, for any $h>0$, the approximation slightly overestimates stability due to the scheme's numerical diffusion (which vanishes as $h\to 0$), resulting in a too small but growing (as $h\to 0$) unstable jamiton region.


\begin{figure}
\begin{subfigure}{.5\textwidth}
  \centering\captionsetup{width=.85\linewidth}%
  \includegraphics[width=.99\linewidth]{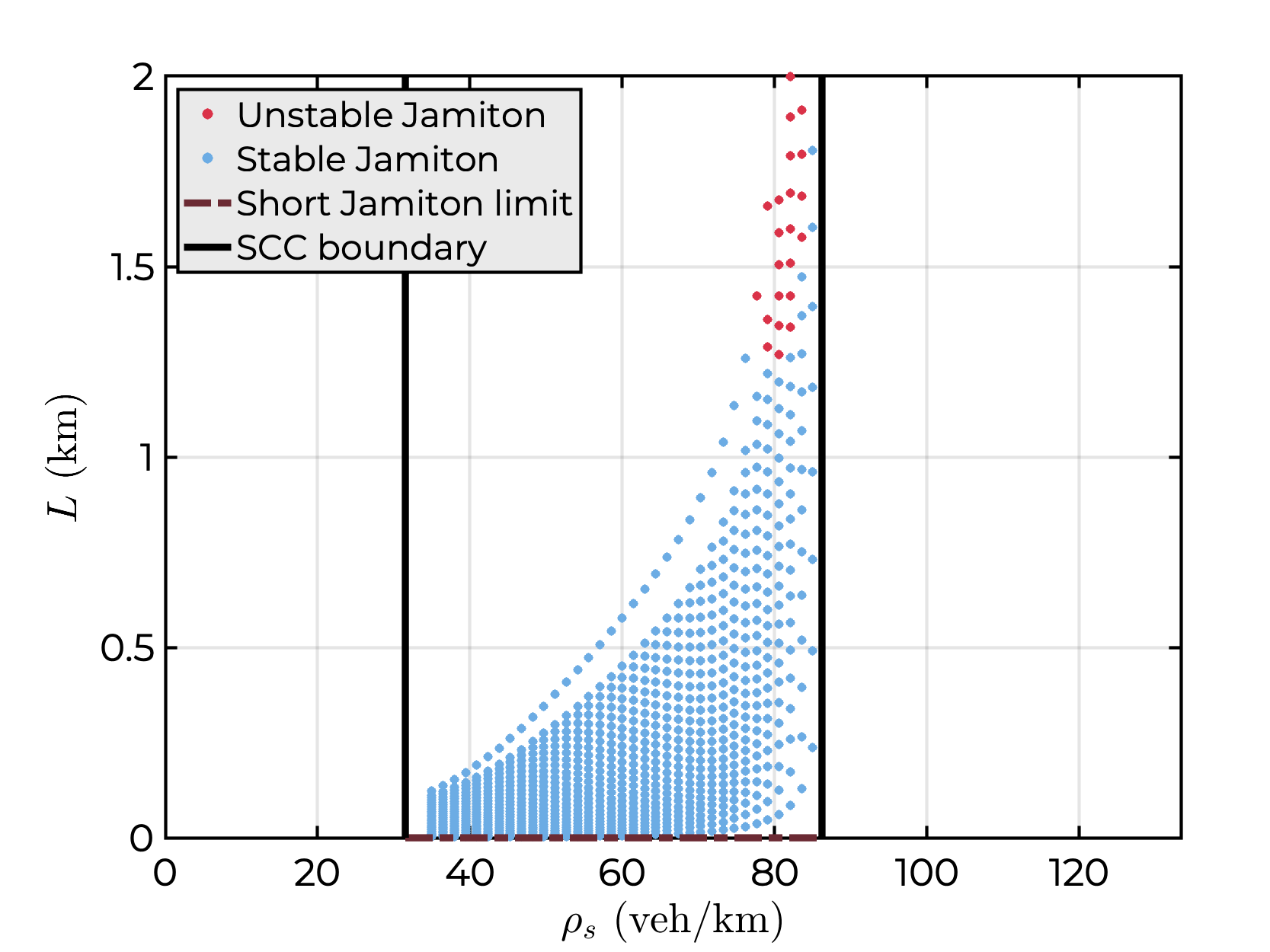}
  \caption{Classification of asymptotic stability in the phase plane $(\rho_\text{S},L)$.}
  \label{subfig:stability_scatter_asymptotic_rhos_vs_L}
\end{subfigure}%
\begin{subfigure}{.5\textwidth}
  \centering\captionsetup{width=.85\linewidth}%
  \includegraphics[width=.99\linewidth]{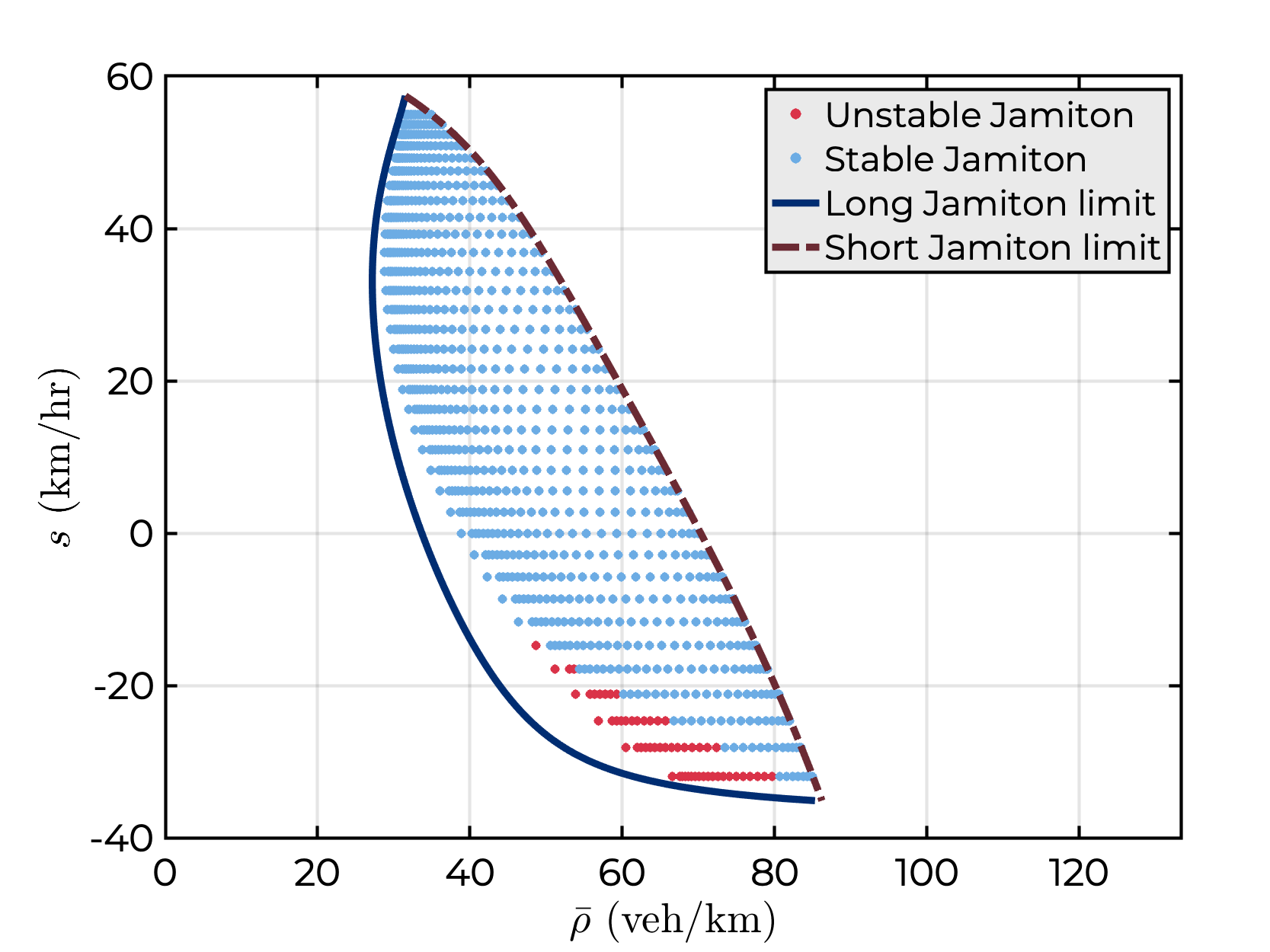}
  \caption{Classification of asymptotic stability in the phase plane $(\bar{\rho},s)$.}
  \label{subfig:stability_scatter_asymptotic_rhobar_vs_s}
\end{subfigure}
\caption{Classification of 980 jamitons into asymptotically stable vs.~unstable, where asymptotic stability is given by $\rho(M) < 1$ (and instability by $\rho(M) > 1$). Here, for each jamiton, $M$ is the one-step update matrix that comes from a discretization with 8000 grid points. Note that the criterion used here can only detect ``splitting'' instabilities.}
\label{fig:stability_scatter_asymptotic}
\end{figure}

Figure~\ref{fig:stability_scatter_asymptotic} displays the results. It shows the classification of the same jamitons as in Fig.~\ref{fig:stability_scatter} into asymptotically stable and unstable using the asymptotic stability criterion: $\rho(M) < 1$ (unstable: $\rho(M) > 1$), where for each jamiton, $M$ is the one step update matrix that comes from a discretization with 8000 grid points. Comparing those results to the nonlinear system results in Fig.~\ref{fig:stability_scatter}, we indeed see that (i)~only the splitting instability (long jamitons) can be captured; and (ii)~the unstable region is underestimated. This last aspect is likely also affected by the fact that asymptotic stability does not account for transient growth effects; which we consider next.

\subsection{Quantitative results: Transient growth}
\label{subsec:transient_growth}
Even if the system \eqref{eq:perturbation_system_PDE_short} is aymptotically stable, small perturbations may be amplified significantly at transient times. Via asymptotic arguments we can argue that the dominant wave amplitude growth mechanism is the growth of the $u$-field as it travels between the sonic point $\xS$ and the right domain boundary $N_0$. The argument (which can be made rigorous via a WKB expansion \cite{BenderOrszag1978}) is as follows.

Consider high frequency solutions of \eqref{eq:perturbation_system_PDE_short}, i.e., solutions that are rapidly varying in space and time. In this situation the behavior is dominated by the left hand side, and we can see that such solutions generally consist of a superposition of two waves: the ``$u$-wave'', dominated by the excitation in $u$, and the ``$q$-wave'', dominated by the excitation in $q$. Consider first the $u$-wave. Then, because $u \gg q$, we can simplify the equations to obtain
\begin{equation*}
\begin{split}
u_t + b_1(x) u_x &\approx a_{11}(x)u\;,\\
q_t + b_2\phantom{(x)} q_x &\approx a_{21}(x)u\;.
\end{split}
\end{equation*}
From this we can see that $q$ is ``slaved'' to $u$ (since the homogeneous part of the solution to the second equation should be considered as belonging to the $q$-wave). A similar argument applies to the $q$-wave; however, the $u$-wave will dominate because $a_{11}>0$ to the right of $\xS$, while $a_{22}<0$.

Hence, neglecting the $q$-wave (and its influence on $u$) we obtain that $u$ evolves (approximately) according to the characteristic equations $\frac{\ud{x}}{\ud{t}} = b_1(x)$ and $\frac{\ud{u}}{\ud{t}} = a_{11}(x)u$. The speed $b_1$ vanishes at $\xS$, but so does the growth rate $a_{11}$, resulting in an overall finite net growth. By the chain rule, the characteristic equations lead to the ODE
$\frac{\ud{u}}{\ud{x}} = \frac{a_{11}(x)}{b_1(x)}u$, with normalized i.c.~$u(\xS) = 1$, to estimate the transient amplification factor $F$. Solving the ODE yields
\begin{equation}
\label{eq:growth_factor}
F = \exp\prn{\int_{\xS}^{N_0}\frac{a_{11}(x)}{b_1(x)}\ud{x}}\;.
\end{equation}
This quantity can be computed via quadrature, using L'H{\^o}pital's rule at/near $\xS$. However, note that the arguments above do not apply across the sonic point, even though the integrand is not singular, because the parameterization of the characteristics by $x$ (i.e.~$\frac{\ud{t}}{\ud{x}} = \frac{1}{b_1(x)}$) implicit in the calculation above breaks down there.

An important fact is that the quantity $F$ can be computed without solving the jamiton ODE. This is achieved by parameterizing the jamiton in terms of $\vS$ and the left shock state $\vv_{N_0} = \vv(N_0)$. Then, because $a_{11}$ and $b_1$ are functions of $x$ only via the jamiton $\vv(x)$, one can apply a change of variables to replace $x$-integration by $\vv$-integration. The Jacobian for the transformation follows from the jamiton ODE \eqref{eq:jamiton_ode}. This yields the formula
\begin{equation*}
F = \exp\prn{ \int_{\vS}^{\vv_{N_0}}
\frac{m_0 h^\ppr(\vv)}{h^\pr(\vv)(h^\pr(\vv)+m_0)} -
\frac{m_0 (h^\pr(\vv)+U^\pr(\vv))}{h^\pr(\vv)(U(\vv)-m_0\vv-s_0)}
\ud{\vv} }\;.
\end{equation*}

\begin{figure}
\begin{subfigure}{.5\textwidth}
  \centering\captionsetup{width=.85\linewidth}%
  \includegraphics[width=.99\linewidth]{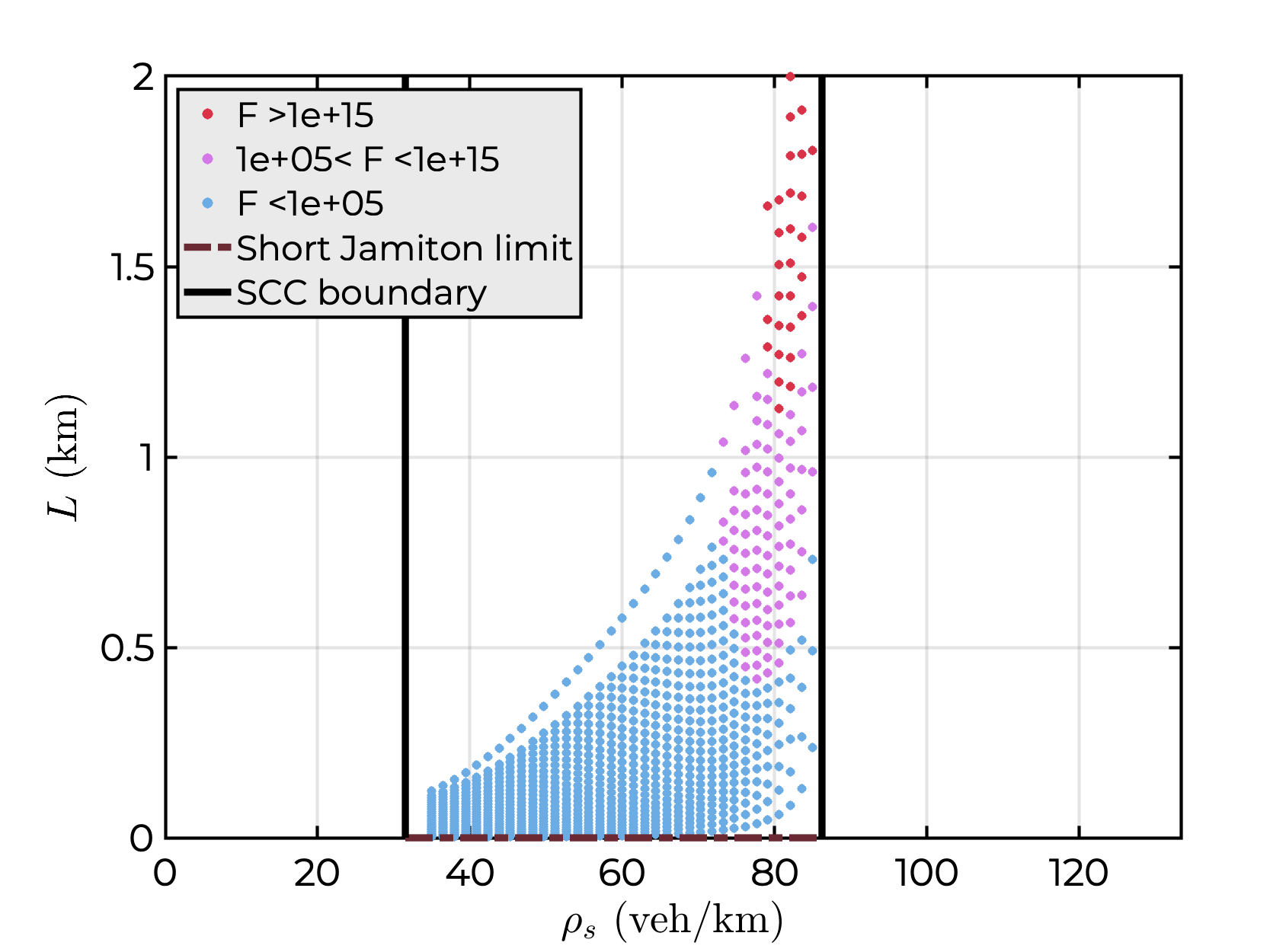}
  \caption{Classification of jamitons according to $F$ in the phase plane $(\rho_\text{S},L)$.}
  \label{subfig:stability_scatter_transient_rhos_vs_L}
\end{subfigure}%
\begin{subfigure}{.5\textwidth}
  \centering\captionsetup{width=.85\linewidth}%
  \includegraphics[width=.99\linewidth]{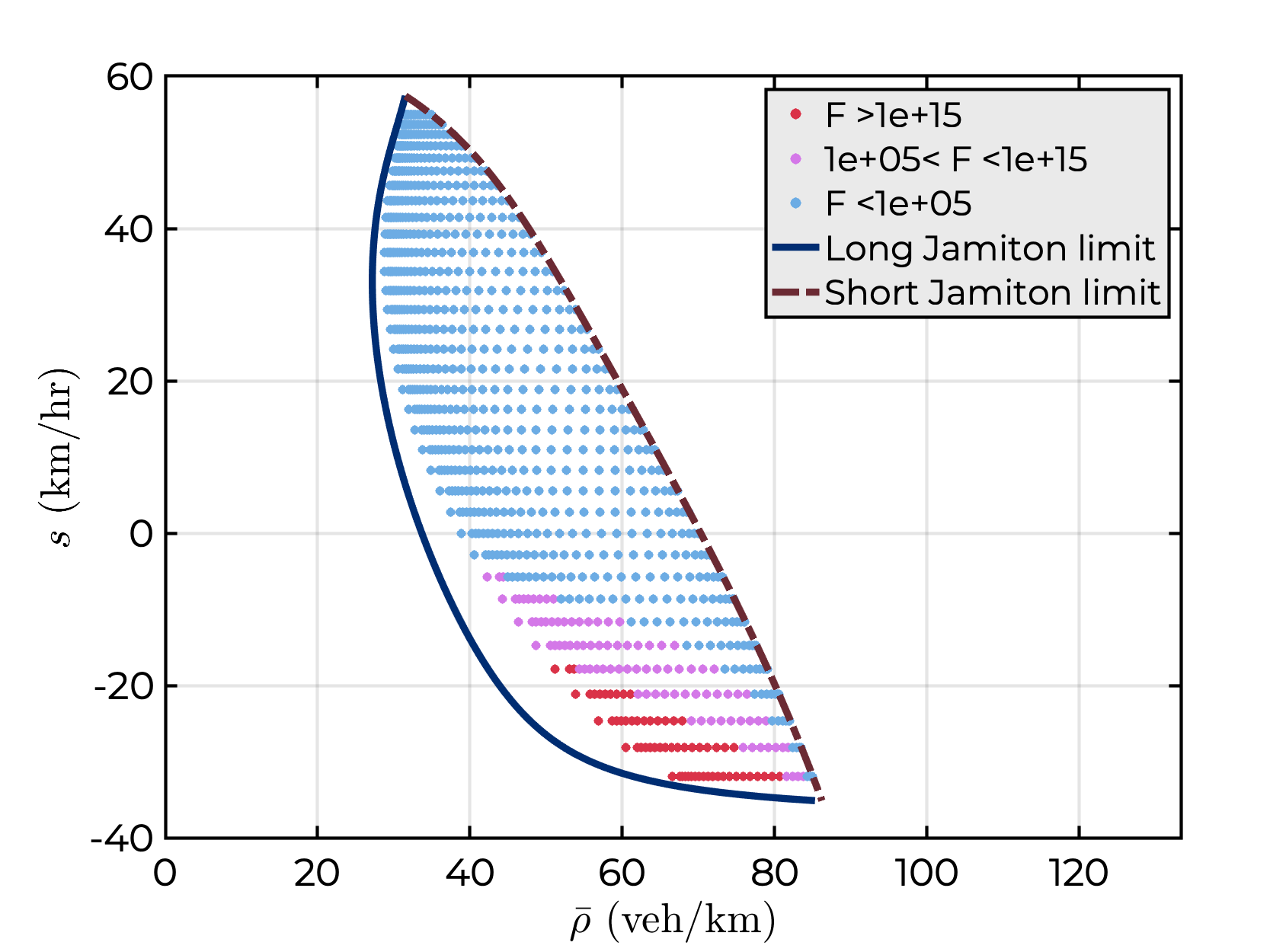}
  \caption{Classification of jamitons according to $F$ in the phase plane $(\bar{\rho},s)$.}
  \label{subfig:stability_scatter_transient_rhobar_vs_s}
\end{subfigure}
\caption{Classification of 980 jamitons according to the transient growth factor \eqref{eq:growth_factor}. Three levels of $F$ are displayed, with the thresholds at $F_1 = 10^5$ and $F_2 = 10^{15}$ to yield: stable if $F<F_1$, moderately unstable if $F_1 < F < F_2$, and unstable if $F_2 < F$.}
\label{fig:stability_scatter_transient}
\end{figure}

Figure~\ref{fig:stability_scatter_transient} shows the stability classification via this criterion for the same jamitons studied in Fig.~\ref{fig:stability_scatter}. As in Fig.~\ref{fig:stability_scatter_asymptotic}, we do not capture merging instabilities. For the splitting instability, we consider two thresholds for the amplification factor: $F_1 = 10^5$ and $F_2 = 10^{15}$. Classifying jamitons below the $10^5$ amplification factor as stable is consistent with the magnitude of noise in the nonlinear computation (\S\ref{subsec:computational_study_results}), which was roughly $10^{-5}$. The results show that the stability boundary in Fig.~\ref{fig:stability_scatter} is not reproduced perfectly, but reasonably well. An interesting advantage of this measure of ``instability'' is that it not just a yes/no criterion, but rather provides a measure of the ``badness'' of the instability. One key missing piece in this criterion is that it does not characterize the ``pumping'' mechanism of perturbations from $q$ into $u$ at/near the sonic point. Hence, we do not know how large the perturbation magnitude really is near $\xS$.

\section{Discussion and Outlook}
\label{sec:conclusions}
The study presented in this paper highlights important structural properties of hyperbolic conservation law systems with relaxation terms, in the regime when the sub-characteristic condition (SCC) is violated. Such PDE are of importance in the macroscopic modeling of vehicular traffic flow (the main focus here), but also for other applications, such as roll waves in open channels \cite{Noble2007a} and circular hydraulic jumps \cite{Kasimov2008}. Furthermore, many of the issues are similar to those that appear in the context of the ZND theory for the stability of Chapman-Jouguet (CJ) detonations \cite{FickettDavis1979}. In fact, jamitons are mathematical analogs of detonation waves \cite{FlynnKasimovNaveRosalesSeibold2009}. While for detonation waves the notion of an SCC does not seem to apply, CJ detonations do have a sonic point, which renders their stability analysis \cite{StewartKasimov2006, ClavinDenet2018} difficult. It is our hope that the relative simplicity of systems such as the ARZ model will provide a route to advance in this challenging topic.

This work provides a pathway to understanding important stability questions for the inhomogeneous ARZ model \eqref{eq:ARZ_model}. In the regime of violated SCC, this model can reproduce the practically relevant \cite{SternCuiDelleMonacheBhadaniBuntingChurchillHamiltonHaulcyPohlmannWuPiccoliSeiboldSprinkleWork2018} phenomena of phantom traffic jams and stop-and-go traffic waves, while preserving the advantages of a macroscopic description (see \S\ref{sec:introduction}). The dynamic stability of jamitons determines which of the many theoretically possible jamiton solutions of the model can/will be selected by the equations' dynamics. The study in \S\ref{sec:computational_study} reveals that short jamitons tend to merge, and long jamitons tend to split, resulting in a middle range of stable jamiton wave lengths. A remarkable aspect about this dynamic selection via (in)stability is that it selects a length scale (range), even though there is no length scale that is explicitly inserted into the model.

The perturbation analysis of jamiton solutions presented here leads to a variable-coefficient linear advection-reaction system whose solutions characterize jamiton stability. As shown in \S\ref{sec:jamiton_stability_analysis}, this system exhibits extremely complex dynamics that may not be suspected at first glance, given its simple fundamental structure. A key reason for those complex dynamics is the zero-transition of one characteristic field, which corresponds to the sonic point in the nonlinear jamiton. While a complete analysis of the behavior of the solutions to the perturbation system remains to be conducted in future work (including a full WKB analysis \cite{BenderOrszag1978}), the qualitative characterization presented herein reveals that there are two key mechanisms for instability that must be considered: first, asymptotic stability that captures the net amplification or decay of infinitesimal perturbation that traverse through periodic jamiton patterns; and second, the transient growth of small perturbations as they travel from near the sonic point down the jamiton profile until they eventually hit the next shock. The quantitative study in \S\ref{sec:jamiton_stability_analysis} reveals that for some jamitons, such transient amplifications may yield noise amplification by many orders of magnitude, which for many practical situations will definitely push the solutions into the fully nonlinear regime.

Based on those stability concepts, two criteria have been developed that are directly verifiable in terms of the model functions rather than requiring nonlinear hyperbolic system simulations. Asymptotic stability reduces to finding the spectral radius of a sparse matrix, which in itself is a non-trivial problem as well, but it is an established standard task in numerical linear algebra. For the transient growth, a proxy criterion has been devised that boils down to a straightforward quadrature of two model functions. When compared with the ``brute force'' nonlinear stability results (\S\ref{sec:computational_study}), those two criteria capture the key qualitative essence of the stability boundary for long jamitons; but to reproduce the precise shape there is still room for improvement via more refined stability criteria.

Mathematically, understanding the solution behavior of relaxation system in which the SCC is violated is a crucial challenge \cite{Liu1987, Li2000, JinKatsoulakis2000}, and this work provides some insight. In addition, the jamiton perturbation system \eqref{eq:perturbation_system_PDE} is full of challenging structure (see \S\ref{subsec:fundamental_challenges}), and this paper provides criteria to characterize its stability properties.

For the key application of traffic flow, the understanding of which jamiton solutions are dynamically stable is a critical step towards determining which models reproduce real-world phenomena best. Moreover, the non-normal structure of the system in \eqref{eq:perturbation_system_PDE}, leading to the transient growth behavior it exhibits (\S\ref{subsec:transient_growth}), has interesting connections to the task of stabilizing traffic flow with a single autonomous vehicle \cite{CuiSeiboldSternWork2017}.

Finally, an obvious extension is to tackle the merging instability as well, and we plan to do so in future work. At least in principle, the methodology of this current work can be extended to include the merging instabilities by allowing multiple shock perturbation.

\section{Acknowledgments}
The authors would like to acknowledge the support by the National Science Foundation. R. R. Rosales and B. Seibold were supported through grants DMS--1719637 and DMS--1719640, respectively. Computations were carried out on Temple University's HPC resources and thus were supported in part by the National Science Foundation through major research instrumentation grant number 1625061.

\bibliographystyle{plain}
\bibliography{references_seibold,references_rrr}

\vspace{1.5em}
\end{document}